\documentclass[useAMS,usenatbib]{mn2e}
\usepackage{graphicx}
\usepackage{times}
\usepackage[english]{babel}
\usepackage{xspace}

\newcommand{\lsim}{\vcenter{\hbox{$<$}\offinterlineskip\hbox{$\sim$}}}
\newcommand{\gsim}{\vcenter{\hbox{$>$}\offinterlineskip\hbox{$\sim$}}}

\def\ngc{{NGC 346}\xspace}

\def\apex{{\it APEX}\xspace}
\def\herschel{{\it Herschel}\xspace}
\def\hst{{\it HST}\xspace}
\def\chandra{{\it Chandra}\xspace}
\def\laboca{{\it Laboca}\xspace}
\def\spitzer{{\it Spitzer}\xspace}

\def\arcsec{\ifmmode ^{\prime\prime} \else $^{\prime\prime}$\xspace \fi}
\def\degree{\ifmmode ^{\circ} \else $^{\circ}$\xspace \fi}
\def\msun{\ifmmode \mathrm{M}_{\odot} \else M$_{\odot}$\xspace \fi}
\def\lsun{\ifmmode \mathrm{L}_{\odot} \else L$_{\odot}$\xspace \fi}
\def\halpha{\ifmmode \mathrm{H}\,\alpha \else H\,$\alpha$\xspace \fi}

\newcommand{\varx}[1]{\ifmmode {#1} \else ${#1}$\xspace \fi}
\def\sfr{SFR\xspace}
\def\sfe{SFE\xspace}
\def\uav{\varx{\langle U\rangle}}

\newcommand{\varxy}[2]{\ifmmode {#1}_\mathrm{#2} \else ${#1}_\mathrm{#2}$\xspace \fi}
\def\gdr{\varxy{r}{gd}}
\def\fpah{\varxy{f}{PAH}}
\def\fracmstar{\varxy{frac}{M\star}}
\def\nstar{\varxy{N}{\star}}

\def\ldust{\varxy{L}{dust}}
\def\mdust{\varxy{M}{dust}}

\def\mstar{\varxy{M}{\star}}
\def\mcat{\varxy{M}{cat}}
\def\mtot{\varxy{M}{tot}}

\def\sigmatwentyfour{\varxy{\Sigma}{24\,{\mu}m}}
\def\sigmatwofifty{\varxy{\Sigma}{250\,{\mu}m}}
\def\sigmadust{\varxy{\Sigma}{dust}}
\def\sigmagas{\varxy{\Sigma}{gas}}
\def\sigmasfr{\varxy{\Sigma}{SFR}}
\def\sigmastar{\varxy{\Sigma}{\star}}
\def\sigmamstar{\varxy{\Sigma}{M_{\star}}}
\def\sigmaallgas{\varxy{\Sigma}{HI+H_{2}}}
\def\sigmaco{\varxy{\Sigma}{CO}}
\def\sigmatir{\varxy{\Sigma}{TIR}}
\def\sigmahalpha{\varxy{\Sigma}{H\alpha}}

\def\deltatsfr{\varxy{{\Delta}t}{SFR}}
\def\tauff{\varxy{{\tau}}{ff}}

\begin{document}
\title[Relation between gas and young stars in NGC 346]{Star-formation
  rates from young-star counts and the structure of the ISM across the
  \ngc/N66 complex in the SMC\thanks{based on observations obtained
    with \hst, \spitzer, \herschel and \apex.}}
\author[S. Hony et al.]
{
    S. Hony$^{1}$\thanks{E-mail:\url{sacha.hony@free.fr}},
    D. A. Gouliermis$^{1,2}$,
    F. Galliano$^{3}$,
    M. Galametz$^{4}$,
    D. Cormier$^{1}$,
    C.-H. R. Chen$^{5}$, \and
    S. Dib$^{6,7}$,
    A. Hughes$^{8,9,2}$,
    R. S. Klessen$^{1,10,11}$,
    J. Roman-Duval$^{12}$,
    L. Smith$^{13}$,
    J.-P. Bernard$^{8,9}$, \and
    C. Bot$^{14}$,
    L. Carlson$^{15}$,
    K. Gordon$^{12,16}$,
    R. Indebetouw$^{15,17}$,
    V. Lebouteiller$^{3}$,
    M.-Y. Lee$^{3}$, \and
    S. C. Madden$^{3}$,
    M. Meixner$^{12,18}$,
    J. Oliveira$^{19}$,
    M. Rubio$^{20}$,
    M. Sauvage$^{3}$,
    R. Wu$^{21}$
\\
\\
\begin{minipage}{1.0\linewidth}
$^{1}$Universit\"{a}t Heidelberg, Zentrum f\"{u}r Astronomie, Institut f\"{u}r Theoretische Astrophysik, Albert-Ueberle-Str. 2, 69120 Heidelberg, Germany
${^2}$Max Planck Institute for Astronomy, K\"{o}nigstuhl 17,  69117 Heidelberg, Germany
$^{3}$Laboratoire AIM, CEA/IRFU/Service d'Astrophysique, Universit\'{e} Paris Diderot, Bat. 709, 91191 Gif-sur-Yvette, France
$^{4}$European Southern Observatory, Karl-Schwarzschild-Str. 2, 85748 Garching-bei-M\"unchen, Germany
$^{5}$Max-Planck-Institut f\"ur Radioastronomie, Auf dem H\"ugel 69, 53121 Bonn, Germany
$^{6}$Niels Bohr International Academy, Niels Bohr Institute, Blegdamsvej 17, 2100, Copenhagen, Denmark
$^{7}$Centre for Star and Planet Formation, University of Copenhagen, {\O}ster Voldgade 5-7, 1350, Copenhagen, Denmark
$^{8}$CNRS, IRAP, 9 Av. Colonel Roche, BP 44346, 31028 Toulouse Cedex 4, France
$^{9}$Universit\'e de Toulouse, UPS-OMP, IRAP, 31028 Toulouse Cedex 4, France
$^{10}$Department of Astronomy and Astrophysics, University of California, 1156 High Street, Santa Cruz, CA 95064, USA
$^{11}$Kavli Institute for Particle Astrophysics and Cosmology, Stanford University, SLAC National Accelerator Laboratory, Menlo Park, CA 94025, USA
$^{12}$Space Telescope Science Institute, 3700 San Martin Drive,
Baltimore, MD 21218, USA
$^{13}$Space Telescope Science Institute and European Space Agency, 3700 San Martin Drive,
Baltimore, MD 21218, USA
$^{14}$Observatoire astronomique de Strasbourg, Universit\'e de Strasbourg, CNRS, UMR 7550, 11 rue de l'Universit\'e, 67000 Strasbourg, France
$^{15}$Department of Astronomy, University of Virginia, PO Box 3818, Charlottesville, VA 22903, USA
$^{16}$Sterrenkundig Observatorium, Universiteit Gent, Gent, Belgium
$^{17}$National Radio Astronomical Observatory, Charlottesville, VA 22904, USA
$^{18}$The Johns Hopkins University, Department of Physics and
Astronomy, 366 Bloomberg Center, 3400 N. Charles Street, Baltimore, MD
21218, USA 
$^{19}$School of Physical \& Geographical Sciences, Lennard-Jones 
 Laboratories, Keele University, Staffordshire ST5 5BG, UK
$^{20}$Departamento de Astronomia, Universidad de Chile, Casilla 36-D, Santiago, Chile
$^{21}$Department of Astronomy, Graduate School of Science, The University of Tokyo, Bunkyo-ku, Tokyo 113-0033, Japan
\end{minipage}
}
\date{Accepted 2015 January 15. Received 2015 January 14; in original
  form 2014 October 30}
\pagerange{\pageref{firstpage}--\pageref{lastpage}} \pubyear{XXX}
\maketitle
\label{firstpage}
\begin{abstract}
  The rate at which interstellar gas is converted into stars, and its
  dependence on environment, is one of the pillars on which our
  understanding of the visible Universe is build. We present a
  comparison of the surface density of young stars (\sigmastar) and
  dust surface density (\sigmadust) across \ngc(N66) in 115
  independent pixels of 6$\times$6\,pc$^{2}$. We find a correlation
  between \sigmastar and \sigmadust with a considerable scatter. A
  power law fit to the data yields a steep relation with an exponent
  of 2.6$\pm$0.2. We convert \sigmadust to gas surface density
  (\sigmagas) and \sigmastar to star formation rate (\sfr) surface
  densities (\sigmasfr), using simple assumptions for the gas-to-dust
  mass ratio and the duration of star formation. The derived total
  \sfr (4$\pm$1$\cdot$10$^{-3}$ \msun\,yr$^{-1}$) is consistent with
  \sfr estimated from the \halpha emission integrated over the \halpha
  nebula. On small scales the \sigmasfr derived using \halpha
  systematically underestimates the count-based \sigmasfr, by up to a
  factor of 10. This is due to ionizing photons escaping the area,
  where the stars are counted. We find that individual 36\,pc$^2$
  pixels fall systematically above integrated disc-galaxies in the
  Schmidt-Kennicutt diagram by on average a factor of $\sim$7. The
  \ngc average SFR over a larger area (90\,pc radius) lies closer to
  the relation but remains high by a factor of $\sim$3. The fraction
  of the total mass (gas plus young stars) locked in young stars is
  systematically high ($\sim$10 per cent) within the central 15\,pc
  and systematically lower outside (2 per cent), which we interpret as
  variations in star formation efficiency. The inner 15\,pc is
  dominated by young stars belonging to a centrally condensed cluster,
  while the outer parts are dominated by a dispersed population.
  Therefore, the observed trend could reflect a change of star
  formation efficiency between clustered and non-clustered
  star-formation.
\end{abstract}


\section{Introduction}
\label{sec:intro}
\begin{figure*}
  \includegraphics[clip,width=\textwidth]{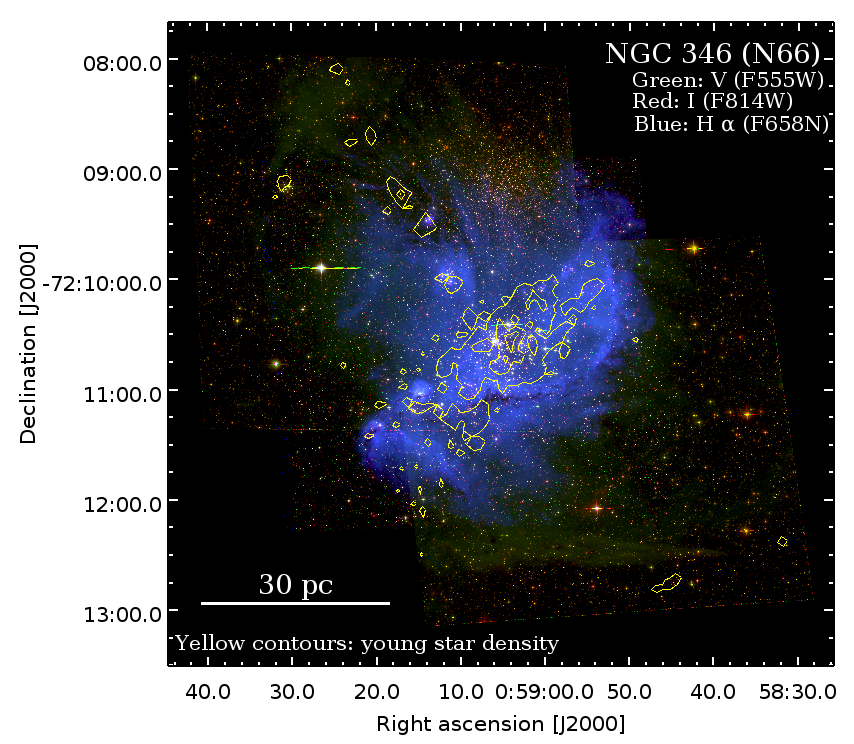}
  \caption{False colour composite image (green: optical, red:
    infrared, blue \halpha) of \ngc based on \hst imaging \citep[PI
    Nota,][]{2006ApJ...640L..29N}. The contours of the, 2\,arcsec
    resolution, density map (\sigmastar) of the young stars are shown
    in yellow. The contour levels correspond to 3, 10 and 20
    stars\,pc$^{-2}$. The maximum value of \sigmastar is 45
    stars\,pc$^{-2}$. }
  \label{fig:hst}
\end{figure*}
It is generally accepted that the local star-formation rate (\sfr)
scales with the local gas density of the interstellar medium (ISM).
This wisdom is captured in the Schmidt-Kennicutt relation
\citep[SK-relation, see for a review][]{2012ARA&A..50..531K} which
connects the observed gas-surface density (\sigmagas) along various
sight-lines with the observed surface density of \sfr (\sigmasfr).
Even though there are multiple incarnations of the Schmidt-Kennicutt
diagram, depending on which tracers are used to measure both \sigmagas
and \sigmasfr, and whether to consider only the dense (molecular) gas
or include the atomic phase, the emerging picture is clear: the higher
the ISM column density, the higher the star-formation rate. A key
feature of the SK-relation is its power law behaviour over a large
dynamic range of gas column-densities and star-formation rates. The
slope of the relation depends on whether one measures only the dense
gas (exponent $\sim$1) or also the atomic gas (exponent $\sim$1.4).

The physical interpretation of this interdependence, i.e. how to
derive the efficiency with which interstellar gas is converted into
stars, is much less self-evident, especially on small spatial
scales ($<$1\,kpc). This is due to the following:
{\it i)} The `star-formation rate' measures stars that have already
formed, while the gas column density relates to (the capacity for)
future star formation. On small scales, with low numbers of stars, one
can no longer expect to measure a time-averaged efficiency as is the
case when averaging over a large number of star forming sites.
{\it ii)} Dynamical simulations \citep[e.g.][]{1998MNRAS.301..759K}
and observations \citep[e.g.][]{2004ASPC..323..171G} show that young
stars may be displaced, by several pc per Myr, from the location where
they were born, and thus removed from the reservoir of gas from which
they formed
{\it iii)} The large scale averages include non-star-forming parts of
the ISM and the relative proportion of this `inactive' gas will
change when zooming in on smaller scales.
{\it iv)} Many of the used \sfr-tracers (e.g. \halpha or
far-ultraviolet (FUV) emission) are sensitive only to high-mass stars,
which represent only a fraction of the stellar mass assembled in the
star-formation event and to derive the total \sfr one needs to
extrapolate the mass-function to lower masses. Since the high-mass
slope of the initial mass function (IMF) is very steep -- very few of
the formed stars are high-mass -- the corrections are large and
sensitive to the assumed shape of the IMF.
Some or all of these effects are probably the reason why the studies
of star formation rates in nearby individual molecular clouds
\citep{2009ApJS..181..321E,2010ApJS..188..313W,2010ApJ...723.1019H,2011ApJ...739...84G}
find very different star formation rates at a given gas column density
compared to global averages.

The aim of this paper is to circumvent some of these complications by
studying the correlation between newly formed stars and the gas
reservoir (\sigmasfr, and \sigmagas) in many small beams ($\sim$6\,pc)
which cover an entire star forming complex. We investigate, thus, the
small scale relation between young stars and gas reservoir while
covering a larger region. We use the most direct way of determining
\sigmasfr, i.e. by counting the newly assembled stars. At the same
time we employ the `dust method' to determine the ISM column density,
i.e. using the infrared (IR) to submillimetre (submm) dust continuum
emission to derive dust surface densities. This method has the
advantage of not being sensitive to the state of the gas (ionized,
neutral or molecular) as long as dust and gas are well mixed. This
insensitivity is important because we will use a catalogue of low- and
intermediate-mass pre-main sequence (PMS) stars plus upper-main
sequence (UMS) stars, i.e. O and B type stars, as the basis for our
star counts. Due to the evolutionary stage of the PMS stars -- a few
Myr after the onset of star formation -- the parental molecular
complex may have been subject to significant photo-dissociation and
using a molecular gas tracer (such as CO) may lead to an incomplete
view of the total gas surface densities.

The object of choice is the star-forming complex related to \ngc, also
often referred to as N66. \ngc is the most prominent star-formation
region in the {Small Magellanic Cloud} (SMC). In order to assess the
\sigmasfr, an unbiased census of the young stellar population is
required. This is available for \ngc thanks to \hst observations
\citep[][see
Sect.~\ref{sec:young-stell-catal}]{2006ApJ...640L..29N,2006ApJS..166..549G}.
We have access to sensitive IR to submm photometry from \spitzer (PI
Gordon), \herschel (PI Meixner) and \apex/\laboca (PI Hony) in order
to derive reliable dust surface densities. The proximity (60\,kpc) of
the SMC in combination with the size of the star forming complex
($\sim$100$\times$100\,pc$^{2}$) enables us to `zoom in'
($\sim$6\,pc linear scale) and study local variations at the angular
resolution of the available submm observations.

Moreover, the entire complex encompasses a range of environments. In
terms of the ISM, the region exhibits a wide range of surface
brightness of hot dust (24~$\mu$m), polycyclic aromatic hydrocarbon
(PAH; 8~$\mu$m) and molecular gas
\citep[CO(J=2-1),][]{2000A&A...359.1139R}. In terms of the stellar
content, there are zones which are cluster dominated as well as
regions where the young stars are more dispersed
\citep{2014MNRAS.439.3775G}. This allows us to examine how the
environment influences the relation between star formation and the
available ISM reservoir. The main questions we address are:
\begin{itemize}
\item How well do young stars and the densest ISM track each other?
\item Do star counts yield the same \sigmasfr as indirect tracers?
\item Does the efficiency of star formation vary between clustered
  and more extended spatial distributions?
\item What is the effect of averaging over different scales on the
  derived parameters in the position in the SK-diagram?
\end{itemize}
Additionally, we examine whether there may be an effect of the reduced
metal abundance of the SMC \citep[0.2 solar,][]{1998A&A...330..285H} on
the star formation efficiency.
\begin{table*}
  \caption{Values and symbols used throughout this paper.}
  \label{tab:symbols}    
  \begin{tabular}{l|l l l}
    \hline
    \hline
    Quantity & Symbol & Value & Comments/Refs \\
    \hline
    SMC distance & $d_\mathrm{SMC}$ & 60$\pm$3 kpc & \citet{2003MNRAS.339..157H} \\
    Detected young stars & $N_\mathrm{star}$ & 5150 & \citet{2006ApJS..166..549G}, see Sect.~\ref{sec:young-stell-catal}\\
    Total young stellar mass & \mtot& 2.2$\cdot$10$^4$\,\msun & \citet{2008AJ....135..173S}, see Sect.~\ref{sec:young-stell-catal}\\
    Mass per catalogue source & \mcat$^{a}$ & 4.3\,\msun & = $\mtot\,N_\mathrm{star}^{-1}$, see Sect.~\ref{sec:young-stell-catal} \\
    SF duration & \deltatsfr & 5$\cdot$10$^{6}$\,yr& \citet{2006AnA...456.1131M}, see Sect.~\ref{sec:young-stell-catal} \\
    Gas-to-dust mass ratio & \gdr & 1250 & Sect.~\ref{sec:conv-dust-surf} \\
    \hline
    Derived Quantity \\
    \hline
    Young star surface density & \sigmastar & & from star catalogue\\
    Stellar mass surf. dens. & \sigmamstar & & $=\sigmastar\cdot\mcat$  \\
    \sfr surf. dens. & \sigmasfr & & $=\sigmamstar\,\deltatsfr^{-1}$  \\
    Dust surf. dens. & \sigmadust  & & from SED fitting \\
    Gas surf. dens. & \sigmagas & & $=\sigmadust\cdot\gdr$ \\
    Stellar mass fraction & \fracmstar & & $=\sigmamstar\,(\sigmamstar+\sigmagas)^{-1}$ \\
    \hline
    \hline
  \end{tabular}
  \newline
  {$^{a}$This mass is \emph{not} the mean mass of the \hst detected
    sources but the mass each source represents after correcting for
    completeness. The mean mass of the young stars in the \hst
    catalogue is $\sim$2\,\msun.}
\end{table*}

The paper is organised as follows. Previous work on deriving \sfr from
individual young sources is reviewed in Sect.~\ref{sec:previous-work}.
In Sect.~\ref{sec:observations} we present the catalogue of young
stars and the IR to submm data-sets that we use for our analysis. We
further describe the data treatment to derive independent surface
density measurements in 20$\times$20\,arcsec$^{2}$
(6$\times$6\,pc$^{2}$) beams (Sect.~\ref{sec:beam-matched-data}). Our
methodology to obtain gas surface densities, using SED fitting, is
described in Sect.~\ref{sec:sed-fitting} and \ref{sec:conv-dust-surf}.
The \sfr derived from our young star counts is compared to indirect
\sfr tracers in Sect.~\ref{sec:sfr-map}. In
Sect.~\ref{sec:comparison-gas-stars} we compare the derived \sigmagas
and \sigmasfr. Possible biases that may affect the observed value of
\sigmasfr/\sigmagas are discussed in Sect.~\ref{sec:possible-biases}.
We conclude in Sect.~\ref{sec:disc--concl}.
   
\section{Previous work}
\label{sec:previous-work}
Most of the work done on comparing ISM column densities and
star-formation rates based on individual objects has concentrated on
nearby (within one kpc from the sun) Milky Way star-forming sites
\citep{2009ApJS..181..321E,2010ApJ...724..687L,2010ApJ...723.1019H,2010ApJS..188..313W,2011ApJ...739...84G}.
These studies have consistently shown that when one considers the
immediate dense environment (on a few parcec scales) surrounding the
newly formed stars, the derived \sfr is higher than predicted by the
galaxy averaged SK-relation (by up to a factor of 54) for a given ISM
column density.

\citet{2009ApJS..181..321E,2010ApJ...724..687L} present the total
star-formation rates based on young stellar object (YSO) candidates,
identified from their mid-IR excess using \spitzer and compare those
to molecular cloud masses derived using extinction maps for $\sim$10
nearby molecular clouds. These studies show that there is a large
variety of star formation efficiency (\sfe) among those clouds.
\citet{2010ApJ...724..687L} show that some of the observed spread in
\sfe disappears when considering only the mass of gas above an
extinction threshold of A$_K$ $>$ 0.8 mag. They interpret this as an
indication for an density threshold below which star-formation does
not occur. A similar conclusion is reached by
\citet{2010ApJ...723.1019H} who study the \sigmasfr versus \sigmagas
within those clouds using YSOs. The typical spatial scale over which
they compare these quantities is from a few tenths of a parsec to a
few parsec. Their data show a very sharp decrease in \sigmasfr below
\sigmagas $\approx$10$^2$ \msun\,pc$^{-2}$ which is interpreted as
resulting from a density threshold for star-formation. Note, that such
a threshold, if it exists, cannot be a \emph{universal} threshold.
There are several observations of specific regions whose gas surface
densities exceed the threshold, with much less star formation than
would be expected
\citep[e.g.][]{2013MNRAS.429..987L,2014ApJ...786..140R}. Moreover, the
existence of a column density-threshold for star formation in a
turbulent ISM is not supported by theoretical models
\citep{2014MNRAS.444.2396C}.
  
A different picture emerges from the analysis presented by
\citet{2011ApJ...739...84G}. These authors present scatter diagrams of
surface density of from YSOs versus \sigmagas (from extinction) for
eight molecular clouds within 1\,kpc from the sun. They find a steep
power law correlation (with indices between 1.4 and 2.7, with an
outlier at 3.8) with significant scatter and no indication for a
density threshold. They argue that the data are consistent with a \sfr
which is proportional to $\sigmagas^2$. Note that the different
conclusion reached by the latter authors is based, in part, on the way
the surface density of YSOs is calculated in the low surface-density
regime as they use the nth-nearest-neighbour centred on the detected
YSOs. We further note that, given the steep dependence on \sigmasfr
and the low number counts, distinguishing between a threshold and this
power law is difficult and almost a matter of semantics.

The work by \citet{2011ApJ...739...84G} is the Galactic study that can
be compared most directly to the \ngc analysis we present here,
because their \sigmagas maps cover several tens of parsecs around the
molecular clouds and each cloud harbours a significant number ($>$100)
of YSOs. Compared to the work we present here, these previous studies
have the advantage of proximity and as a result, they are able to
examine the star-formation within individual molecular clouds on
spatial scales of $<$1\,pc while the best angular resolution we can
attain in the SMC in the ISM tracers is $\sim$6\,pc.

Recent \spitzer surveys of the Large and Small Magellanic Clouds (LMC
and SMC) made it possible to use resolved massive YSOs to determine
current {\sfr}s outside of the Galaxy. Unlike the consistently higher
{\sfr}s of small-size clouds reported in the aforementioned Galactic
studies, \citet{2010ApJ...721.1206C} showed that current \sfr
determined from massive YSOs in individual giant molecular clouds in
H~{\sc ii} complexes in the LMC are distributed above and below the SK
relation, depending on the evolutionary states of the giant molecular
cloud. The scatter in \sfr diminishes and follow the SK relation when
properties are averaged over the entire H~{\sc ii} complexes of
150$-$200\,pc. This is in agreement with the findings of
\citet{2014MNRAS.439.3239K}, who attribute the scatter on smaller
scales to incomplete statistical sampling of independent star-forming
regions. Furthermore, the commonly used \sfr tracers such as \halpha
and 24\,$\mu$m luminosities would under-estimate the amount of star
formation in regions not actively producing O-type stars, such as the
LMC's largest molecular ridge and the Magellanic Bridge. They show
that using \halpha or 24\,$\mu$m as \sfr tracers these points are
outliers on the SK-relation while they follow the SK-relation when
using \sfr determined from YSO counts
\citep{2008AJ....136.1442I,2014ApJ...785..162C}.

The work presented here differs from these previous studies by the
sources we use to trace the star formation (PMS+UMS stars versus YSOs)
and the way we derive the ISM column densities (dust emission versus
extinction). This has the following \mbox{(dis-)advantages}:
\begin{itemize}
\item Rich stellar catalogue: \hst has provided accurate photometry
  for more than 5\,000 stellar sources covering the
  pre-main-sequence \citep{2006ApJS..166..549G}.
\item Measured mass function and age: ages and the mass function
  have been determined for the UMS+PMS stellar catalogue
  \citep{2008AJ....135..173S}. This means we can rely on the
  observed mass function, which covers a significant
  (0.8$-$40\,\msun) part of the entire mass range, and the observed
  duration of the star-formation event to derive the \sfr. The YSOs
  studies assume a typical age and mass for a YSO, both of which are
  uncertain \citep[e.g.][]{2012RAA....12....1P}, 
  which could introduce a large \emph{systematic} uncertainty.
\item Large coverage: the IR and submm observations that we use to
  measure dust column densities are very sensitive and most cover the
  entire SMC (Sect.~\ref{sec:hersch-spitz-data}). As a result, we have
  access to an unbiased probe of the dust surface density out to large
  distances and low column densities. This is important to assess the
  behaviour of the \sfr at low \sigmagas, where significant
  disagreement exists in the literature (see above).
\end{itemize}
However, since we are using stars that are no longer embedded, the
sources in our catalogue are more evolved ($<$5\,Myr,
Sect.~\ref{sec:young-stell-catal}) than YSOs ($<$2\,Myr) and have had
more time to displace (or be displaced from) their parental gas
reservoir. Moreover, since the whole region has to be more evolved for
it to host so many PMS stars, the radiation from the hot, young stars
has had more time to affect the surrounding medium.
  
\section{Observations}
\label{sec:observations}
\subsection{Young stellar catalogue}
\label{sec:young-stell-catal}
Imaging with the Hubble Space Telescope (\hst) of star-forming regions
in the Magellanic Clouds has provided access to exceptionally rich
samples of newly-born stars down to the sub-solar regime over large
areas of the sky. Our catalogue of young stars in \ngc, one of the
largest star-forming complexes in the SMC, is among the richest
Magellanic Clouds stellar ensembles ever collected with \hst. Three
pointings (with significant offsets), centred on \ngc were observed
with the Wide-Field Channel (WFC) of ACS covering an area of about
5$\times$5\,arcmin$^{2}$ ($\sim$88$\times$88\,pc$^2$). The images were
obtained within the HST Program GO-10248 \citep[PI
Nota,][]{2006ApJ...640L..29N} in the filters F555W and F814W
(equivalent to standard V and I bands). Detailed descriptions of the
data reduction and photometry are given in
\citet{2006ApJS..166..549G}. The photometry reveals more than 98\,000
stars at various evolutionary stages, and its 50\% completeness limit
is $m_{\rm 555} \simeq$\,27 \citep{2006ApJS..166..549G}.

The sample of young stellar populations in \ngc consists of the
low-mass pre-main-sequence (PMS) stars, which are identified from
their faint-red positions on the colour--magnitude diagram (CMD). See
\citet{2012SSRv..169....1G} for a review on low-mass PMS stars in the
Magellanic Clouds. This sample is complemented by the UMS stars
(selected with colours $m_{\rm 555} - m_{\rm 814} \leq 0.0$ and
magnitudes 12\,$\sim$\,$m_{\rm 555}$\,$\lsim$\,17), including more
than 30 early-type stars \citep[see
also][]{masseyetal89,2006AnA...456.1131M}. The masses of the selected
UMS stars are such that they would have evolved off the main sequence
if they were older than the PMS population. Thus, the catalogue we use
in this analysis contains 5\,150 young (PMS and UMS) stars in the mass
range between $\sim$\,0.5\,\msun and $\sim$\,40 \msun. The location of
the young stars within the complex is indicated in Fig.~\ref{fig:hst}
with yellow contours. This image shows the very prominent \halpha
emission associated with the central cluster and the presence of
several secondary clusters of young stars to the North which cause
little \halpha emission.

It is hard to get accurate age and mass determinations of the
\emph{individual} PMS stars based on their position in the CMD
\citep{jeffries12}. An isochronal median age around $\sim$\,3\,Myr has
been defined for the \emph{ensemble} of PMS stars in \ngc on the basis
of comparison with evolutionary models
\citep{2008AJ....135..173S,2011AJ....141...31C}.
\citet{2006AnA...456.1131M} have spectroscopically studied the
high-mass stars towards \ngc. They find ages between 0 and 5\,Myr for
the high-mass stars that belong to the star-formation complex. This
age range is fully compatible with the position and spread of the
low-mass PMS stars in the CMD. Since our objective is to measure the
\emph{average} \sfr counting \emph{all} stars that have formed during
the star-formation event, we use 5\,Myr as the star-formation duration
(\deltatsfr). Thus, \deltatsfr refers to the duration of the event
\emph{not} to the ages of the individual stars. Therefore, this
approach is conceptually different from the YSO-counting studies
\citep[e.g.][]{2010ApJ...724..687L,2010ApJ...723.1019H} in which ages
of the individual objects have been assumed.

We construct the entire mass function of the UMS+PMS catalogue by
extrapolating the mass function for masses above 0.8\,\msun derived by
\citet[][Fig 3 bottom panel]{2008AJ....135..173S} from the same
observational material. Their mass function is completeness corrected
for crowding and photometric sensitivity but not for interstellar
extinction. Over the measured range the mass function has a slope of
-1.43\footnote{We use the formalism of the mass function --
  log$_{10}$($N$($\Delta$log$_{10}$($m$))) versus log$_{10}$($m$) --
  as given by \citet{1986FCPh...11....1S}. In this definition the
  Salpeter mass function has a slope of -1.35 and the Kroupa mass
  function has a slope of -1.3 above 0.5 \msun}. We extrapolate to
lower masses using a Kroupa mass function with a slope of -0.3 between
0.08 and 0.5 \msun and a slope of 0.7 between 0.01 and 0.08 \msun
\citep{2001MNRAS.322..231K}. The total stellar mass formed in the last
5\,Myr, obtained by integrating this mass function and multiplying by
the survey area (88$\times$88\,pc$^2$) equals $\sim$2.24$\cdot$10$^4$
\msun. Since we have 5\,150 stars in our catalogue, each source
represents on average a mass of 4.3\,\msun (\mcat). The mass in
directly detected stars (\mstar $>$ 0.8\,\msun) is
1.2$\cdot$10$^4$\,\msun, i.e. 50 per cent of the total. The main
parameters used in our analysis are summarised in
Tab.~\ref{tab:symbols}.

The total star formation rate, averaged over the last 5\,Myr, of the
complex calculated in this way is 4$\pm$1$\cdot$10$^{-3}$
\msun\,yr$^{-1}$, where the uncertainty takes into account possible
star formation outside of the field-of-view of \hst (20 per cent) and the
extrapolation of the observed mass function to lower masses (20 per cent) but
not the systematic uncertainty due the assumed duration of the star
formation event. This value is consistent with the results of
\citet{2007ApJ...669..327S} who find a \sfr$>$3.2$\cdot$10$^{-3}$
\msun\,yr$^{-1}$ over the last 1\,Myr using \spitzer identified YSO
candidates.

For the rest of the analysis we assume that the mass function and the
duration of star formation are invariant across the star forming
complex. There are some indications of variations in the slope of the
mass function in \ngc \citep{2008AJ....135..173S,2011AJ....141...31C}.
The general trend is a steeping of the slope with increasing distance
to the central cluster, i.e. a larger fraction of high-mass stars
towards the central cluster, consistent with a certain degree of mass
segregation in the cluster. However, such variations have little
impact on our results which are based on star counts, while luminosity
based \sfr tracers (e.g. \halpha) are very sensitive to the number of
high-mass stars and thus the high-mass slope of the mass function. As
an illustration of this invariance consider the standard Kroupa mass
function \citep{2001MNRAS.322..231K} with a high-mass power law slope
of -1.3. When we vary this slope between -1.0 and -2.0 -- the observed
range of slopes in \ngc \citep{2008AJ....135..173S} -- the total mass
per number of stars above 0.8\,\msun (i.e. \mcat) varies by only
$\sim$20 per cent. For the same variations in slope the fraction of
mass locked in the high-mass stars -- those powering the \halpha
emission (\mstar$>$8\,\msun) -- varies by 1500 per cent!

\subsection{\herschel and \spitzer data}
\label{sec:hersch-spitz-data}
\begin{figure}
  \includegraphics[clip,width=8.8cm]{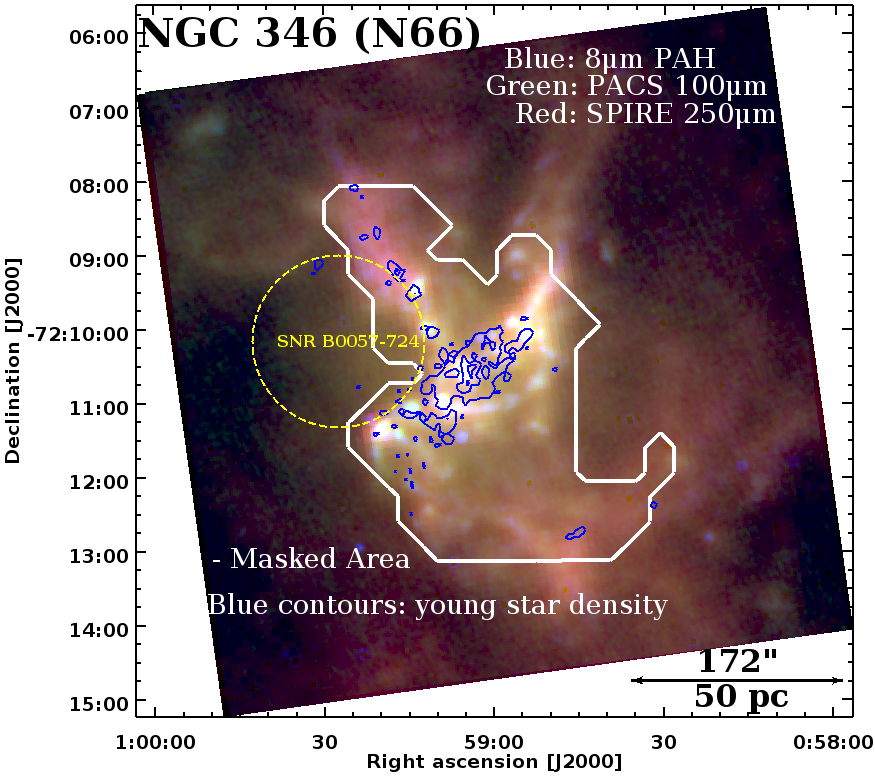}
  \caption{False colour image showing an overview of the large
    environment of \ngc in key ISM tracers. We show 8~$\mu$m emission
    indicative of the location of the most prominent PDRs (blue; IRAC4
    emission minus the stellar continuum estimated from IRAC2), the
    warmer dust traced by PACS 100 (green) and the cooler dust (SPIRE
    250, red). Each waveband is shown at its native resolution. The
    blue contours show the stellar density (see Fig.~\ref{fig:hst}).
    The general observed trend is a decreasing dust temperature with
    distance to the most rich cluster (yellow to red colour gradient)
    and strong PDRs near (but not peaking on) the stellar
    concentrations. The region where both dust surface density,
    including \laboca measurements, and stellar densities are well
    defined is indicated by the white line (see
    Sect.~\ref{sec:beam-matched-data} for details). }
  \label{fig:overview}
\end{figure}
We use the mid-IR to submm photometric maps that have been
obtained with \spitzer \citep[SAGE-SMC:][]{2011AJ....142..102G} and
\herschel \citep[HERITAGE:][]{2013AJ....146...62M}. Within these two
large programs the entire SMC has been observed at IRAC (3.6, 4.5 5.8
and 8.0\,$\mu$m), MIPS (24, 70, and 160\,$\mu$m), PACS (100 and
160\,$\mu$m) and SPIRE (250, 350 and 500\,$\mu$m) wavelengths. The
sensitivity of these data sets is such that for the environment of the
prominent star forming region \ngc the photometric uncertainty is
dominated by the absolute flux calibration (see also
Sect.~\ref{sec:beam-matched-data}). An overview of \ngc in some key
ISM tracers is shown in Fig.~\ref{fig:overview}. In the following
analysis we mainly use the IRAC, MIPS-24, MIPS-70, PACS-100, PACS-160
and SPIRE-250 data, because these sufficiently cover the spectral
energy distribution (SED) to derive accurate dust surface densities at
optimal angular resolution ($\sim$20\,arcsec). The SPIRE 350~$\mu$m
and 500~$\mu$m data have poorer angular resolution. The only exception
is the dust masses derived within large circular apertures
(Tab.~\ref{tab:averages}) which uses the SPIRE-350\,$\mu$m data to
replace the \laboca because the latter is not observed over such a
large area.

\subsection{\laboca 870~$\mu$m imaging data}
\label{sec:laboca-870-mum}
\begin{figure}
  \includegraphics[clip,width=8.8cm]{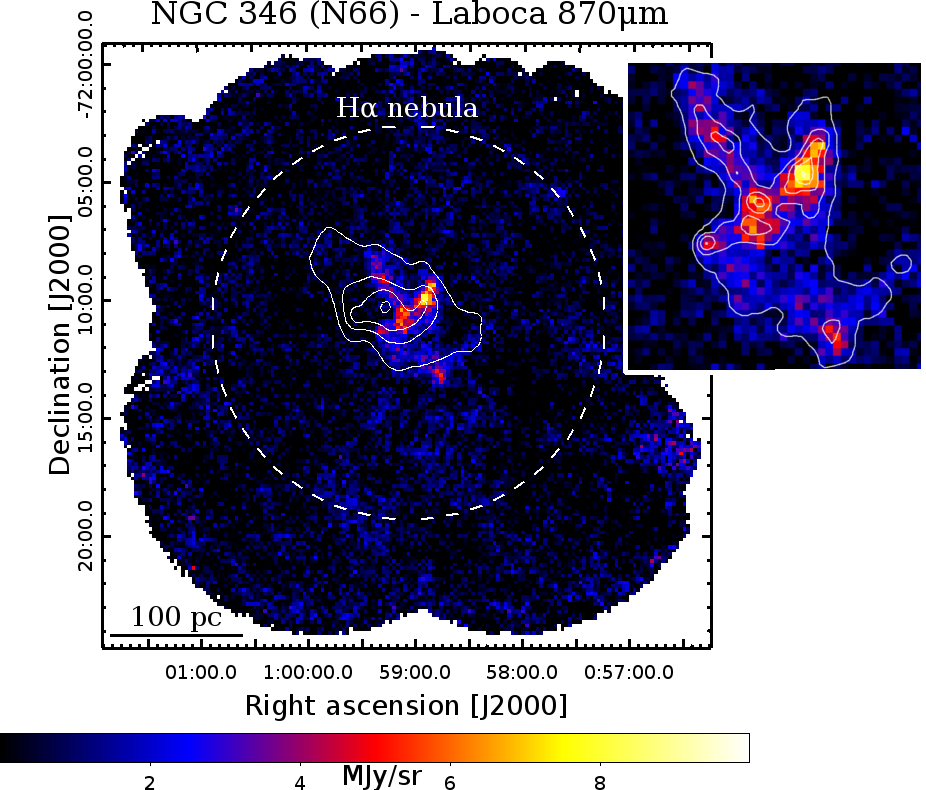}
  \caption{\laboca 870~$\mu$m map of \ngc. The contours show the 36
    cm radio continuum emission from SUMSS
    \citep{2003MNRAS.342.1117M}. The 3 and 6\,cm radio continuum
    images (not shown) exhibit the same morphology
    \citep{2010AJ....140.1511D}. The possible corresponding
    contribution of free-free emission at 870\,$\mu$m has been
    filtered-out by the median filtering applied in the data
    reduction process. The dashed circle indicates the approximate
    extent of the \halpha nebula. We show a close-up of the
    brightest region in the inset, in which the 250\,$\mu$m dust
    emission contours, in perfect correspondence with the \laboca
    map, are shown. }
  \label{fig:laboca}
\end{figure}
To constrain the emission by the coolest dust, we use \laboca data at
870\,$\mu$m, instead of the SPIRE 350 data. This data set allows us to
obtain measurements at $\sim$20\,arcsec resolution which is
unattainable with \herschel/SPIRE 350 and 500. \laboca is a
multi-channel bolometer array for continuum observations at 870$\mu$m
installed on the Atacama Pathfinder EXperiment (APEX) telescope in
Chile. It has a field-of-view of 11.4\,arcmin and a PSF
full-width-at-half-max of 19.2$\pm$0.3\,arcsec. Observations of the
\ngc region with \laboca were carried out in December 2008 and April,
May and September 2009 (Program ID: O-081.F-9329A-2008 – PI: Hony).
The mapping was done with a raster of spiral pattern in order to fully
sample the final map (size of the region mapped:
25$\times$25\,arcmin$^2$). Data are reduced with the BOlometer Array
Analysis Software \citep[BoA][]{boa_manual}. The scans are reduced
individually. They are calibrated using the observations of planets
(Mars, Neptune, Uranus, Venus) and secondary calibrators (G10.62,
G5.89, HLTAU, PKS0506.61, PMNJ0210-5101, PMNJ0303-6211, PMNJ0450-8100,
N2071IR and VY-CMa). We determine zenith opacities using a linear
combination of the opacity determined using sky-dips and that computed
from the precipitable water vapor\footnote{Tabulated sky opacities are
  provided at
  \url{http://www.apex-telescope.org/bolometer/laboca/calibration}}.
We remove dead or noisy channels as well as stationary points or data
taken at fast scanning velocity or above an acceleration threshold. We
subtract the correlated noise, apply steps of median noise removal,
baseline correction and despiking. The reduced scans are then combined
into a final map. We isolate the pixels above a given signal-to-noise
ratio to define a `source model' that is subtracted from the
time-ordered data. We rerun the reduction pipeline and add the source
model to the time-ordered data to obtain a new map. The process is
iterated until the result map converges (5 iterations were applied for
\ngc)\footnote{More details on the techniques are provided in
  \citet{2013MNRAS.431.1596G}.}. The final map is shown in
Fig.~\ref{fig:laboca}. Within the region of interest there are three
main emission features. There is an emission peak towards the central
cluster, a direct indication that this sight-line is dust rich. We
detect a brighter emission peak towards the north-west of the main
cluster and a plume of 870~$\mu$m emission which follows the string of
stellar clustering towards the north-west. This plume coincides with
the brightest CO(j=2-1) emission detected in the region
\cite{2000A&A...359.1139R}. Due to the median filtering that is
applied to remove the atmospheric and telescope signal, any spatially
very extended emission has been removed as well. This explains why we
do not detect any free-free emission from the ionized gas which should
be present in this region (see Fig.~\ref{fig:laboca})

\subsection{Beam matched data}
\label{sec:beam-matched-data}
In our analysis, we use maps with a resolution of 20\,arcsec (6\,pc at
the distance of the SMC). All maps with intrinsic spatial resolutions
better than this resolution, including the stellar catalogue, have
been convolved to the this effective spatial resolution. \laboca,
MIPS\,70 and SPIRE\,250 maps have not been convolved since they
already have an intrinsic spatial resolution very close to 20\,arcsec.
We use the convolution kernels available from Karl Gordons web
page\footnote{\url{http://dirty.as.arizona.edu/~kgordon/mips/conv_psfs/conv_psfs.html}}
for convolving to the SPIRE 250 beams. For the stellar density map --
which has an exceedingly good intrinsic angular resolution we have
directly convolved the 2\,arcsec stellar density map with the SPIRE
250~$\mu$m beam. After convolution, all maps have been re-projected on
the same pixels scheme with pixels the size of
20$\times$20\,arcsec$^{2}$ in order to have spatially independent
measurements and to allow pixel-by-pixel SED extraction and comparison
between the stellar densities and derived dust surface densities. We
use the program {\it swarp} from
AstroMatic\footnote{\url{http://www.astromatic.net}} for calculating
the re-projected maps, as it is in our experience the most accurate
tool in conserving the mean surface brightness. This is particularly
true when re-projecting to a pixel scheme with very different pixel
sizes, which is our case \footnote{The {\sc IDL}/astrolib tool {\it
    hastrom}, often used for the same purpose, performs poorly in this
  case.}.

In order to create reliable random noise maps that represent well
the noise in the original maps and the effects of convolution and
re-projection we apply the following procedure:
\begin{itemize}
\item The amplitude of the noise in each map has been estimated by
  measuring the median absolute deviation from the median (MAD). This
  value is a more robust measure of the noise in a map than the
  standard deviation in the presence of real source emission. The
  estimated standard deviation equals 1.4825*MAD.
\item We generate a pure noise map at the original resolution using
  the measured standard deviation.
\item We generate a projected noise map convolving and re-projecting
  this noise map in the exact same manner as we do the science map.
\item We measure the amplitude of the noise in the projected noise map
  as the standard deviation of the fitted Gaussian function to the
  histogram of pixel values.
\end{itemize}

These output noise levels have been compared to the formal calibration
uncertainties (expressed as percentages of the observed surface
brightness levels). For all maps except the \laboca map, the absolute
calibration is dominant throughout the region of interest. In the case
of the \laboca map, we retain for each pixel in the projected noise
map the larger value of the flux calibration uncertainty and the
random noise value. These uncertainties should not be added
(quadratically) because the \laboca flux calibration uncertainty
reflects the scatter in the measured calibrator fluxes and is thus, in
essence, also a random noise term.

Pixels outside the area surveyed by \hst and pixels where the \laboca
measurement is less than 2 $\sigma$ are masked. There are 115 valid,
independent 20$\times$20\,arcsec$^{2}$ pixels (covering about
$\sim$4000 pc$^2$) in the following pixel-by-pixel analysis. Their
outline is shown in Fig.~\ref{fig:overview}. All maps at the other
wavelengths have signal-to-noise ratios larger than 2 -- in most cases
much larger -- in these pixels.
  
\section{Dust, gas and \sfr surface densities}
\label{sec:dust-gas-column}
\subsection{SED fitting}
\label{sec:sed-fitting}
\begin{figure}
  \includegraphics[clip,width=8.8cm]{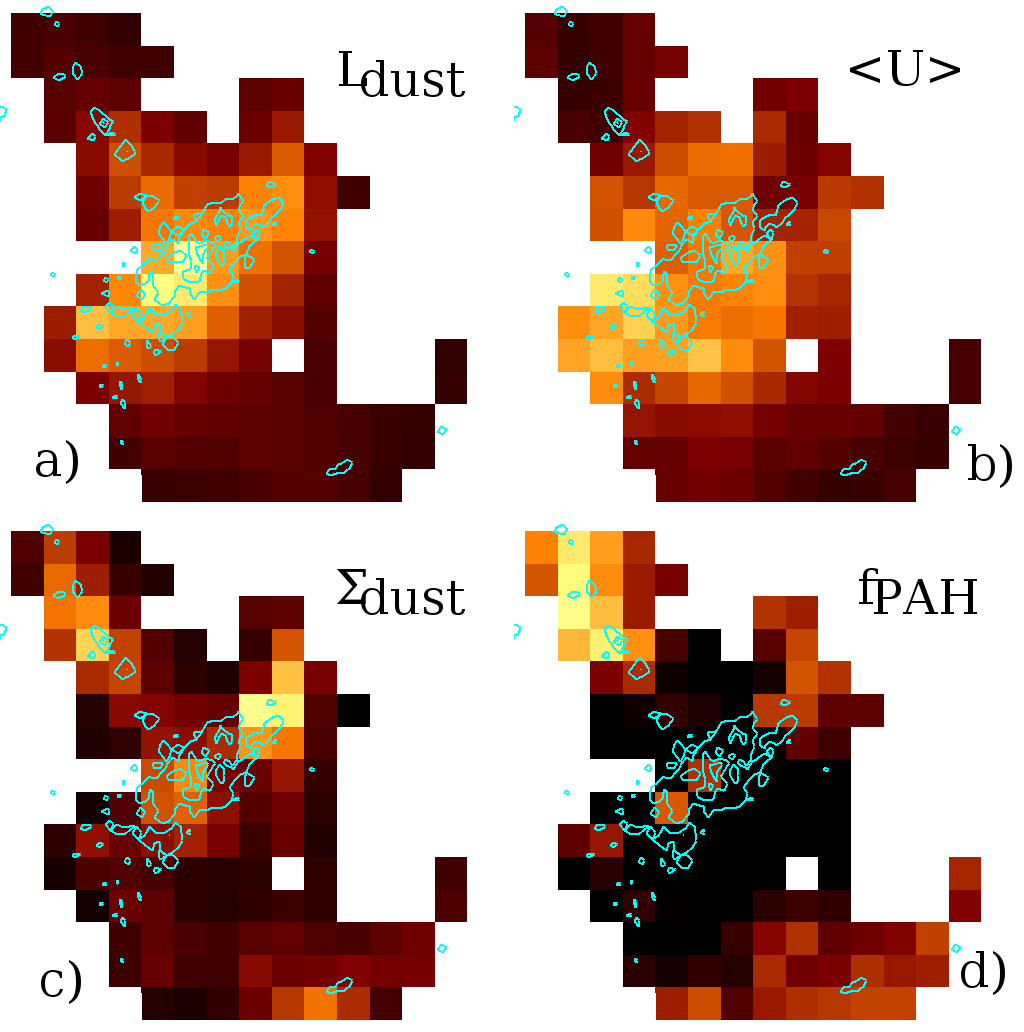}
  \caption{Maps of the output parameters of the dust SED fitting. The
    region shown corresponds to the white contour in
    Fig.~\ref{fig:overview}. For reference we show the surface density
    map of young stars in contours (see Fig.~\ref{fig:hst}). We show
    the luminosity of the dust emission ranging from 150 to 1300 \lsun
    (a), the average radiation field as experienced by the dust
    ranging between 6 and 45 G$_0$ (b), dust surface density ranging
    from 0.05 to 0.3 \msun\,pc$^{-2}$ (c) and the mass of PAHs
    relative to the total dust mass normalised to the Galactic value
    ranging from 0 to 0.3 (d).}
  \label{fig:fitmaps}
\end{figure}
In order to derive dust surface densities, we fit the IR to submm dust
continuum SED for each independent pixel in the region. The dust
emission fitting is performed with the \citet{galliano11} SED model.
This model accounts for variations of dust excitation rates along the
line-of-sight. Following the prescription of \citet{dale01}, it
assumes that the distribution of dust mass, $M_d$, is parametrized by
the starlight intensity, $U$, as a power law:
${\rm d}M_d/{\rm d}U \propto U^{-\alpha}$, between $U_{min}$ and
$U_{min}+\Delta U$. The parameter $U$ is the stellar flux received by
the dust and is normalised to the solar neighbourhood value
\citep{mathis83}. The shape of the SED (in particular, both its peak
wavelength and its width) constrain the parameters of this power law.

For a given starlight intensity, $U$, the dust emissivity is derived
from the \citet[][dust family: BARE-GR-S]{zubko04} model, designed to
fit the diffuse Galactic emission. However, we replace graphite by
amorphous carbon \citep[][ACAR]{zubko96}. Indeed, \citet{galliano11}
showed that the standard Galactic grain properties applied to \spitzer
and \herschel observations of the {LMC} lead to violation of elemental
abundance constraints, while the modified composition (the `AC model')
gives realistic gas-to-dust mass ratios. In summary, we make the
assumption that grains in the {SMC} will have similar optical
properties as in the {LMC}.

The choice of a set of grain emissivities is tainted by systematic
uncertainties. As demonstrated by \citet{2011A&A...536A..88G}, the
derived dust mass change by a factor of 2 to 3 depending on the
optical properties. However, it is also shown that the change is
essentially a simple scaling of the dust mass, and do not
significantly affect the trends. Besides, if the origin of the
difference in emissivity between the Milky Way and the LMC is an
evolutionary effect, linked to metallicity
\citep[cf.][]{2011A&A...536A..17P}, it is likely that using LMC
emissivities puts us closer to the reality than simply assuming
Galactic properties.

The fit of the observed SED, with the previously described dust model,
provides constraints on: the dust mass, $M_d$; the parameters of the
power law distribution of starlight intensities, $U_{min}$, $\Delta U$
and $\alpha$, which we summarise by quoting the average starlight
intensity, \uav; and the PAH mass fraction, normalised to the Galactic
value, \fpah. In order to estimate the final uncertainties on these
derived dust parameters, we perform Monte-Carlo iterations, following
the method described in \citet{galliano11}, taking into account the
correlation between calibration uncertainties. The median
uncertainties on the derived parameters ($\sigma_x$/$x$) are 6 per cent for
\ldust, and 19 per cent for \sigmadust and \uav. The median standard
deviation on \fpah is 0.02.

Maps of the best fit output parameters are shown in
Fig.~\ref{fig:fitmaps}. One of the key results from the SED fitting
which is evident from these maps is the clear distinction between the
parameters which trace the heating of the dust (panels a and b) and
the surface density of dust (panel c). Clearly the dust luminosity in
the very active star forming region \ngc (panel a) traces closely the
radiation field. This is in stark contrast to other (more usual)
observations of more diffuse regions where the incident radiation
field (and dust temperature) does not vary much and \ldust traces the
dust surface density. Panel d shows that the PAH abundance close to
the main cluster is so low that their emission is undetectable in the
broadband photometry. This low abundance of PAH is in perfect
agreement with the findings of \citet{1990A&A...237..296S} who deduced
an increased differential depletion of PAHs towards \ngc based on IRAS
photometry. Similar paucities of PAHs towards other high-mass star
forming region in the Magellanic Clouds have been reported
\citep[e.g.][]{2010ApJ...715..701S,2010A&A...518L..76H}.

\subsection{Conversion of dust surface density to gas surface
  density}
\label{sec:conv-dust-surf}
\begin{table}
  \caption{Total dust mass, gas consumption time-scale and integrated
    SFE as a function of  the area considered in the analysis. The
    first row corresponds to the area included 
    in the pixel-by-pixel analysis. A radius of 52~pc is equivalent
    to the area covered by the \hst observations. 
    The 90\,pc radius matches the prominent SPIRE 250 and 350~$\mu$m waveband
    emission, while the largest radius probes the full extent of the
    \halpha nebula around \ngc. The gas consumption time-scale
    ($\tau_{gas}$) is defined as the M$_{gas}$/\sfr and the
    integrated star-formation efficiency (SFE$_{10 Myr}$) is
    calculated as the percentage of gas that will have been converted into
    stars after 10\,Myr, assuming that the \sfr remains constant. We
    use 10\,Myr as the limiting age, when supernovae might disrupt the
  ISM and halt the star formation.}
  \label{tab:averages}
  \begin{center}
    \begin{tabular}{l|r r r}
      \hline
      \hline
      area/radius & \mdust & $\tau_{gas}$ & SFE$_{10 Myr}$ \\
                  & [\msun] & [10$^8$ yr] & [\%]\\
      \hline
      mask &  498 $\pm$ 18\% &  1.7 & 5.8 \\
      52 pc &  900 $\pm$ 19\% &  2.5 & 4.0 \\
      90 pc & 1616 $\pm$ 19\% &  4.5 & 2.2 \\
      145 pc & 2741 $\pm$ 22\% &  7.6 & 1.3 \\
      \hline
      \hline
    \end{tabular}
  \end{center}
\end{table}
We assume that gas and dust are well mixed along each line-of-sight in
which case the gas surface density (\sigmagas) is proportional to the
dust surface density (\sigmadust) and the proportionality factor is
the gas-to-dust mass ratio (\gdr). The dust masses derived by SED
fitting methods vary significantly and systematically
\citep[e.g.][]{2011A&A...536A..88G,2014ApJ...797...85G} depending on
the applied method and assumed dust optical properties. It is
therefore important to use a value of \gdr consistent with the used
SED fitting methodology, i.e. the value of \gdr that gives the correct
gas mass given the derived dust mass. We have calibrated the \gdr
using the available gas tracers on a large scale. Within a 500\,arcsec
radius centred on \ngc we find an atomic gas mass of
2.7$\cdot$10$^6$\,\msun using the Parkes+ATCA HI map
\citep{2003ApJS..148..473K} and a spin temperature of 60K
\citep{2008AJ....136..919B}. The molecular gas mass determined from
CO(J=1-0) within the same aperture is 7.4$\cdot$10$^4$ \msun when
using a CO-to-H$_2$ column density conversion factor ($X_{CO}$) of
10$^{21}$ [cm$^{-2}$ (K\,km\,s$^{-1}$)$^{-1}$]
\citep{2013ARA&A..51..207B}. If we use an $X_{CO}$ of 10$^{22}$
[cm$^{-2}$ (K\,km\,s$^{-1}$)$^{-1}$] we find a molecular gas mass of
7.5$\cdot$10$^{5}$ \msun. Such an elevated $X_{CO}$ may be more
appropriate given the harsh radiation field that is prevalent in the
region, which may induce enhanced photo-dissociation of the CO
molecules. Thus, we derive a total gas mass in the range of
2.7$\cdot$10$^{6}$ to 3.4$\cdot$10$^{6}$ \msun. The dust mass derived
using the dust continuum photometry in this aperture and applying the
same SED fitting routine we use for the pixel-by-pixel analysis yields
2.7$\cdot$10$^{3}$ \msun. The derived \gdr ranges from $\sim$1000 to
$\sim$1250. In the following we use a fiducial value for \gdr of 1250,
which is much higher than the Galactic value
\citep[100,][]{2007ApJ...657..810D} due to the lower metallicity of
the SMC. Our value for \gdr of 1250 is lower than the total SMC
integrated value of 1740 from \citet{2014ApJ...797...85G}. The global
value includes a significant amount of atomic gas at low column
density in the outskirts of the SMC which is not detected in dust
emission. This drives up the global \gdr.

Given that the derivation of \gdr is susceptible to significant
\emph{systematic} uncertainties in $X_{CO}$ for the gas surface
densities and dust optical properties for the dust surface densities,
we will discuss the implications of a factor of two variations in the
absolute value of \gdr on our results. Note that, part of the purpose
of this paper is to determine how well the ISM and stars follow each
other on small spatial scales. This part of the analysis does not
depend on the chosen absolute value for \gdr. The comparison of \sfr
and star-formation efficiencies with values derived for other regions
is however sensitive to the chosen value and we will discuss the
influence of our choice on the results we find.

We further assume \gdr to be invariant for the entire region. In
principle, it is possible that \gdr could vary significantly, in
particular in the vicinity of high-mass star-formation sites, where
dust destruction can become important due to supernovae explosions.
\citet{2006MNRAS.367.1379R} have studied the supernovae remnants
within the large environment of \ngc. These authors show that the
region we study here appears largely unaffected by the impact of
supernovae. The closest remnant they identified ({SNR B0057-724}, see
also Fig.~\ref{fig:overview}) lies to the North-East of the main
cluster, outside of the region of interest. The observational data
argue against strong variations of \gdr because they show that the
ratio of dust heating compared to gas heating is very constant
throughout the whole region (see Sect.~\ref{sec:variations-gdr}).

\subsection{Star formation rate surface densities}
\label{sec:sfr-map}
\begin{figure}
  \includegraphics[clip,width=8.8cm]{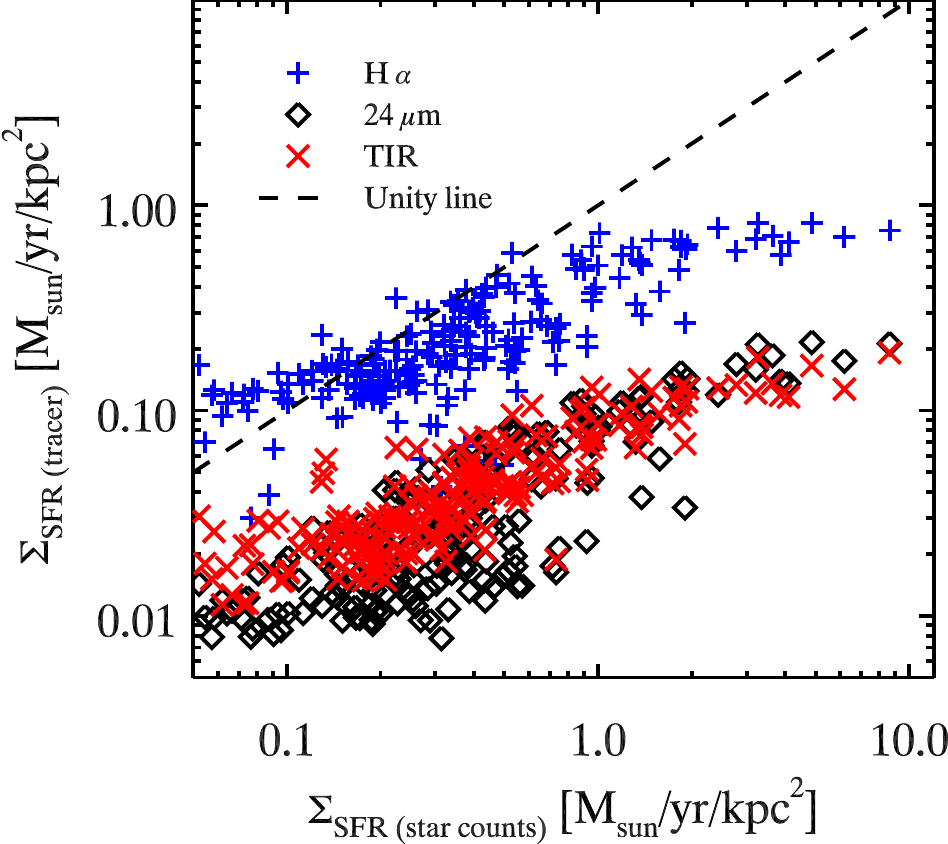}
  \caption{Comparison of derived pixel-by-pixel \sigmasfr using
    different methods. The x-axis shows the \sigmasfr using the
    young-star surface density. In blue we show \sigmasfr calculated
    following \citet[][for the Kroupa IMF]{2011ApJ...741..124H} using
    the continuum subtracted \halpha map form MCELS
    \citep{2000ASPC..221...83S} that was kindly provided by S. Points.
    The red crosses are \sigmasfr estimates using the total infrared
    (TIR) (3$-$1100~$\mu$m) surface brightness following
    \citet{2011ApJ...741..124H} and the black symbols show the
    estimated \sigmasfr based on \spitzer/MIPS 24~$\mu$m
    \citep{2009ApJ...692..556R}. }
  \label{fig:sfr_comparison}
\end{figure}
\begin{figure}
  \includegraphics[clip,width=8cm]{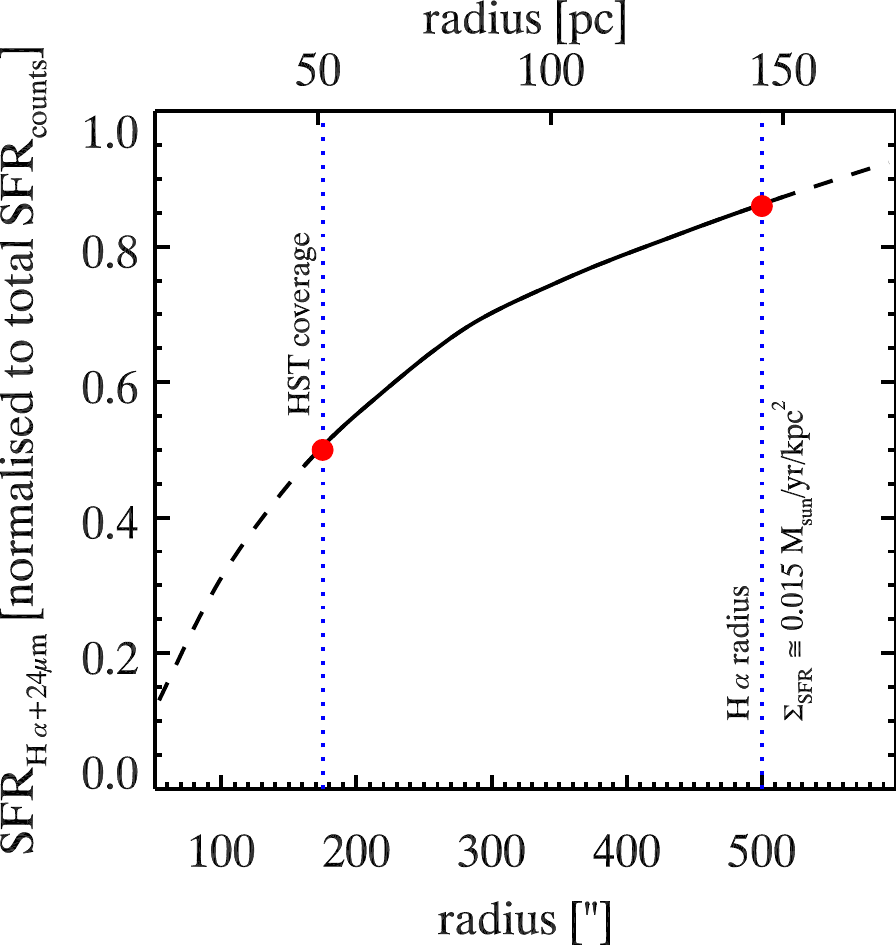}
  \caption{Derived total \sfr for the \ngc complex based on the
    combined emission of \halpha and 24 $\mu$m as function of the
    radius (centred on 14.776\degree;-72.1735\degree) over which the
    flux is integrated. We follow the prescription from
    \citet[][Eqn.~5]{2007ApJ...666..870C}. The \sfr is expressed in
    terms of the fraction of the total \sfr derived from star counts.
    The \halpha and 24\,$\mu$m emission are much more extended than
    the distribution of the young stars. As can be seen, the \sfr
    derived within the radius of the \hst coverage ($\sim$50 pc) is
    about half of the total \sfr. Only when we integrate over the
    complete nebula ($\sim$500\,arcsec), where the \halpha emission
    reaches the diffuse background level do we derive similar \sfr
    values.}
  \label{fig:halpha_sfr_radius}
\end{figure}
The star-formation-rate map has been derived using the observed number
of young stars along the line-of-sight in each pixel of the
beam-matched stellar density map. It is calculated as (see also
Table~\ref{tab:symbols}):
\begin{equation}
  \label{eq:2}
  \mathrm{\sfr} = \frac{\nstar \mcat}{\deltatsfr} = \frac{\nstar \cdot 4.3}{5 \cdot
    10^6} ~~~ [\msun\,\mathrm{yr}^{-1}]
\end{equation}
The star-formation surface density map (\sigmasfr) is obtained by
dividing this map by the area of a pixel in kpc$^2$.

In Fig.~\ref{fig:sfr_comparison}, we compare the derived \sigmasfr
with selected, indirect star-formation rate tracers. Note that we only
compare with those tracers (\sigmahalpha -- calibration from
\citet{2011ApJ...741..124H} for a Kroupa IMF; \sigmatwentyfour --
calibration from \citet{2009ApJ...692..556R} and \sigmatir --
calibration from \citealt{2011ApJ...737...67M}) which are appropriate
for the very recent epoch of star-formation ($<$5\,Myr) and we do not
consider FUV because its calibration assumes a constant \sfr over
$>$10$-$100\,Myr. As can be seen, the \sigmasfr from star counts is
generally higher than those obtained from \halpha or dust emission.
Moreover, the differences are largest where the stellar density (i.e.
\sigmasfr) is highest. Towards the densest cluster the difference
between counting stars and using \halpha is a factor of 10, while the
dust tracers yield a \emph{local} \sigmasfr lower by a factor of 50.
The dust tracers yield much lower values (by a factor of $\sim$5) than
both \halpha due to the reduced dust abundance in the SMC. 

The origin of the weak dust emission in \ngc is not the same as the
weak emission in the Gould-belt star-formation regions reported by
\citet{2013ApJ...765..129V}. These authors have compared the \sfr one
would derive from \spitzer/MIPS 24~$\mu$m measurements or TIR with
\sfr derived from counting YSOs. They find that the IR tracers
underpredict the true star formation rate by two orders of magnitude,
because these regions lack high-mass stars due to incomplete sampling
of the initial mass function. \ngc is a very rich environment and does
not have such a lack of high-mass stars.

The difference between young star counts and the \halpha derived
values is due to the fact that the individual massive young stars in
\ngc produce \halpha (and dust) emission over a much larger surface
than the size of the pixel in which they are counted, significantly
larger than the full \hst field-of-view. Note, that this effect will
cause the tight relation between \sfr (derived from \halpha) and the
gas column density to break down on small ($\lsim$100\,pc) scales,
similar to what is observed in {M33} \citep{2010ApJ...722L.127O}. This
break-down occurs even in the absence of drifting of the stars from
their birth environment or spread induced by evolutionary effects,
simply because the local \halpha (or dust) emission no longer measures
the locally produced UV-radiation field.
Figure~\ref{fig:halpha_sfr_radius} demonstrates how large this
distance is. We show the fraction of the \sfr which is recovered when
integrating the extinction corrected \halpha emission (based on the
combination of \halpha and 24~$\mu$m emission) in apertures of
increasing radius. Within the \hst field-of-view we recover 50 per
cent. When we integrate out to a radius of 500\,arcsec (145\,pc),
where the \halpha emission from the \ngc complex becomes
indistinguishable from the general SMC emission, we recover $\sim$90
per cent. In this calculation we assume that outside the field-of-view
of the \hst observations, the star formation is insignificant. This is
corroborated by the results of \citet{2013ApJ...778...15S} who
identified YSO candidates based on IR photometry of the entire SMC. Of
the sources from their highly-reliable YSO-candidate list which are
associated with \ngc, less than 20 per cent (4/23) are located outside
of the field-of-view of \hst. A very similar result is found when
using the spectroscopically confirmed early-type stars from
\citet{2010AJ....140..416B}. Of the 43 O-type stars from this
catalogue within a 100\,pc radius around \ngc only $\sim$15 per cent
(6) are not within the field-of-view of \hst.

We conclude that the total \sfr as measured from the (predominantly
low-mass) young star counts is consistent with the \sfr measured from
\halpha emission to within $\sim$10 per cent. However, the absorption
of the hard radiation from the hot stars and subsequent re-emission as
either recombination lines or dust continuum occurs far -- at
distances $>$100 pc, in the case of \ngc -- from the actual location
of these stars. See also \citet[]{2012MNRAS.423.2933R} and
\citet{2013ApJ...768..180L} for excellent discussions on how radiation
escaping the local environment affects \halpha and dust emission as
\sfr tracers on small spatial scales. As a consequence, these indirect
tracers which depend on absorption and re-emission, cannot be used for
our present purpose, which is to relate the stellar densities with the
ISM on smaller scales within the star forming complex. Note that most
extra-galactic studies that use \halpha to derived \sfr are probing
much larger regions ($\gsim$kpc) than the scale on which the indirect
tracers are reconciled with the direct measurements ($\sim$150\,pc)
and are thus unaffected by this effect.
  
\section{Comparison of gas and stars}
\label{sec:comparison-gas-stars}
In the following section, we compare the derived \sfr and ISM column
densities. We first concentrate on the global behaviour and how \ngc
compares to other measurements before looking in more detail at the
tendencies within the complex.
\subsection{Global comparison}
\label{sec:global-comparison}
\begin{figure}
  \includegraphics[clip,width=8.8cm]{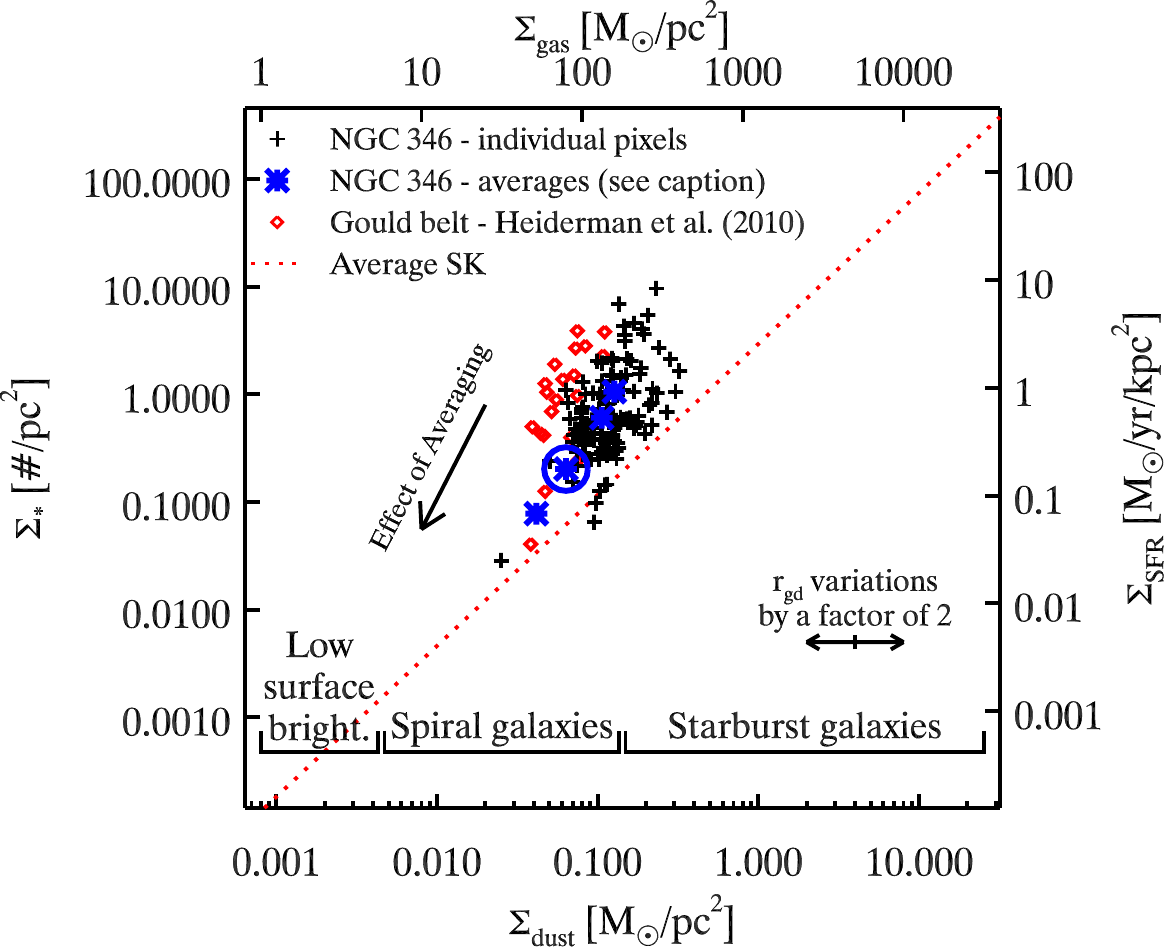}
  \caption{Scatter plot of surface density of young stars versus
    surface density of dust per pixel of 20$\times$20\,arcsec$^{2}$
    (6$\times$6\,pc$^{2}$). The dotted line shows the `standard'
    relation from \citet[][Fig.~11a]{2012ARA&A..50..531K}. The values
    from \citet{2010ApJ...723.1019H}, based on YSO counts and ISM
    column densities in nearby individual molecular clumps, are also
    shown. The values for \ngc (black pluses) lie systematically above
    the `standard' line. The thick blue stars indicate where \ngc is
    located when we average over all pixels in the masked area, and
    subsequently over apertures with radii of 52, 90 and 145\,pc (from
    top-right to bottom left). These radii correspond to the
    approximate size of the HST surveyed area, the prominent far-IR
    emission nebula and the size of the \halpha nebula, respectively.
    The power law exponent of the effect of averaging, i.e the slope
    of the arrow in the figure, is 2.9.}
  \label{fig:Sigma_dust_Sigma_stars_large}
\end{figure}
We show in Fig.~\ref{fig:Sigma_dust_Sigma_stars_large} the stellar
surface density versus the dust surface density for each pixel. These
are the basic observational quantities which do not depend on
assumptions about \mcat, \deltatsfr or \gdr. Using the values in
Table~\ref{tab:symbols} we convert these into \sigmasfr and \sigmagas
(right and top axes). The individual pixels within \ngc lie above the
fiducial SK-relation for \sigmasfr versus \sigmaallgas
\citep[][Fig.~11a]{2012ARA&A..50..531K} by on average a factor of 7.
For reference, we also show the points for individual clumps in nearby
Milky Way star forming regions from \citet{2010ApJ...723.1019H}, which
also lie consistently above the SK-relation. These authors use \sfr
from low-mass YSOs and column densities from the dense gas measured
from extinction maps on a typical scale of a few parsec. 
The fact that all these small-scale measurements lie above the galaxy
averaged values is expected \citep{2014MNRAS.439.3239K}. The offset is
due to the selection biased introduced by studying a region which is
currently, actively forming stars, while the galaxy averaged values
include also regions which are not presently forming stars.
Compared to the study from \citet{2010ApJ...723.1019H}, our approach
has the advantage of using the dust in \emph{emission} to measure the
column densities and that we have access to dust emission covering a
large area surrounding the star forming complex. This allows us to
calculate averaged values over relevant spatial scales, taking into
account also parts of the ISM which are less vigorously forming stars.

The thick blue stars correspond to averages over the entire masked
area, the entire \hst field-of-view, the prominent SPIRE 250~$\mu$m
emission and the \halpha nebula (top-right to bottom left,
respectively, see also Tab.~\ref{tab:averages}). The mean value lies
progressively closer to the SK-relation as we increase the area over
which we calculate the average. It should be noted that the same
shifting to bottom left \emph{must} occur also to the values from
\citet{2010ApJ...723.1019H} if they were to integrate to larger
distance from their young stars (to lower column densities), as the
conditions that prevail in their low-mass star forming regions are
probably more representative for the spiral galaxies at low \sigmasfr
rather than those on the dividing line between the most active spirals
and star burst galaxies. The third star from the top in
Fig.~\ref{fig:Sigma_dust_Sigma_stars_large} -- highlighted by a circle
-- indicates the average over the area where significant cold dust
emission is detected in excess to the general SMC emission. This is
probably the best estimate of the average \sfr and ISM column density
for the complex in its entirety. This value is above the SK-relation
by about a factor of three albeit roughly compatible with the average
relation when taking into account its scatter of 0.3dex.
This factor is consistent with the prediction from
\citet{2014MNRAS.439.3239K} for the size of the region $\sim$300\,pc
diameter over which we do the comparison.
We also indicate in Fig.~\ref{fig:Sigma_dust_Sigma_stars_large}
by how much the \ngc values would be displaced (left-to-right)
relative to the data from \citet{2010ApJ...723.1019H} and the SK line
as a result of a factor of two variations in \gdr. If \gdr were 2500
as opposed to 1250, the average values of \ngc would fall on the SK
line. Such a high value for \gdr would imply that half of the total
gas (more than 80 per cent of the molecular gas) within a 500\,arcsec
(145\,pc) radius around \ngc is molecular without corresponding CO
emission, as a result of photo-dissociation of the CO molecule. Such
an large reservoir of hidden molecular gas, not related to CO, seems
difficult to reconcile with the emission of C$^+$ which traces the
photo-dissociation. \citet{2011A&A...531A..19I} show that the
distribution of the [C {\sc ii}]~158\,$\mu$m line emission closely
resembles the CO distribution.

\subsection{The physical origin of variations in stellar mass
  fraction}
\label{sec:clues-phys-orig}
\begin{figure}
  \includegraphics[clip,width=8.8cm]{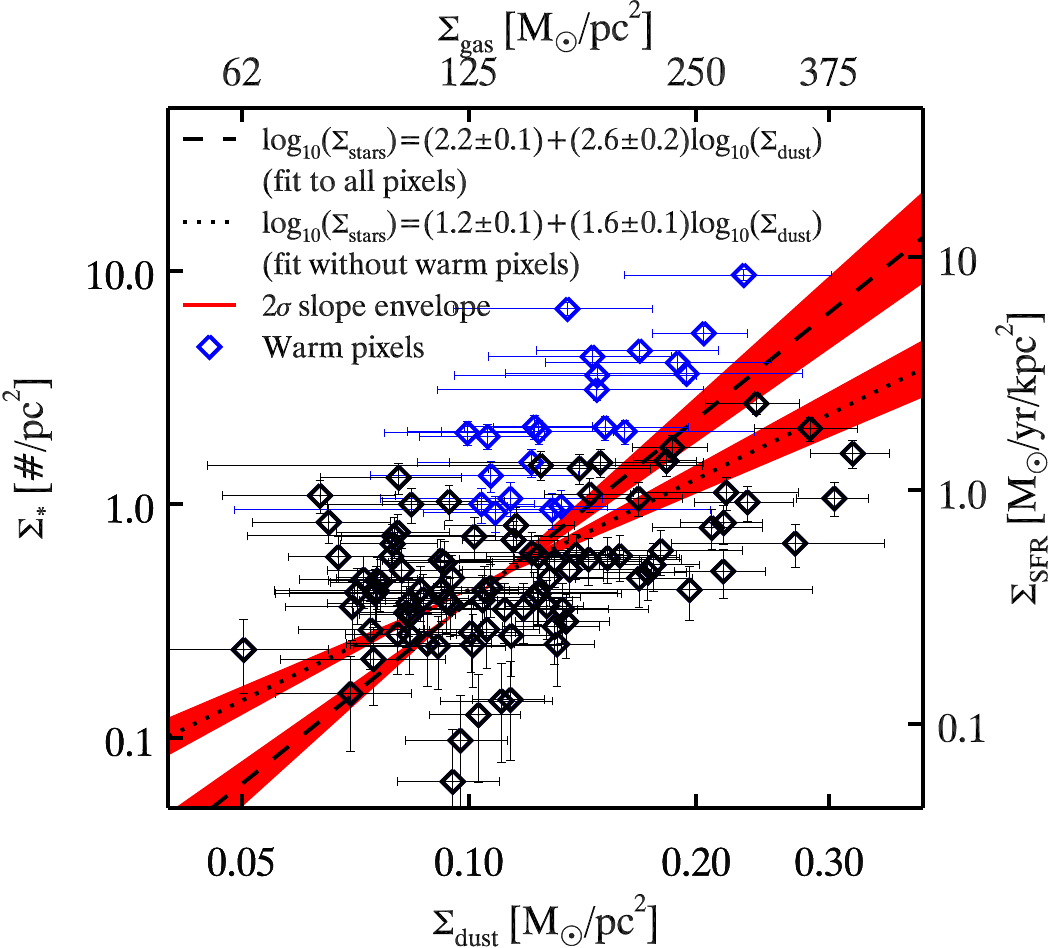}
  \caption{Scatter plot of surface density of young stars versus
    surface density of dust per pixel of 20$\times$20\,arcsec$^{2}$
    (6$\times$6\,pc$^{2}$) zoomed in on the range occupied by \ngc.
    The blue points have a strong warm dust component in their SED
    (\sigmatwentyfour/\sigmatwofifty [F$_\nu$/F$_\nu$] $>$ 0.3).
    Clearly, the lines-of-sight selected using this colour criterion
    occupy a separate region of the diagram. The long and short dashed
    lines are the straight-line fits in logarithmic space to the data
    with or without the warm pixels, respectively. The red areas
    corresponds to the 2$\sigma$ uncertainties of the best-fitting
    slope, while keeping the normalisation fixed. }
  \label{fig:Sigma_dust_Sigma_stars}
\end{figure}
On the spatial scale that we probe here ($\sim$6$\times$6\,pc$^{2}$)
we measure significant scatter in the relation between \sigmagas and
\sigmamstar (Fig.~\ref{fig:Sigma_dust_Sigma_stars_large}). Such
scatter on small scales is expected, due to the stochastic nature of
the local star-formation process, as well as relative motions between
the formed stars and the reservoir from which they formed
\citep{2014MNRAS.439.3239K}. The observed scatter is more easily
assessed in Fig.~\ref{fig:Sigma_dust_Sigma_stars}, where we
concentrate on the values derived for \ngc. This figure also shows the
derived power law for the entire ensemble of sight-lines. We use the
{\sc IDL} routine MPFITEXY\footnote{\url{http://purl.org/mike/mpfitexy}}
\citep{2010MNRAS.409.1330W} which uses MPFIT
\cite{2009ASPC..411..251M} to determine a straight-line fit to the
data (in log-log space) with uncertainties in both variables. The
best fit value for the exponent is 2.6$\pm$0.2, i.e. the relation is
much steeper that the canonical SK-relation (exponent: $\sim$1.4).
This very steep increase of \sigmasfr with \sigmagas is similar to the
large exponents found by \citet{2011ApJ...739...84G}.

\begin{figure}
  \includegraphics[clip,width=8.8cm]{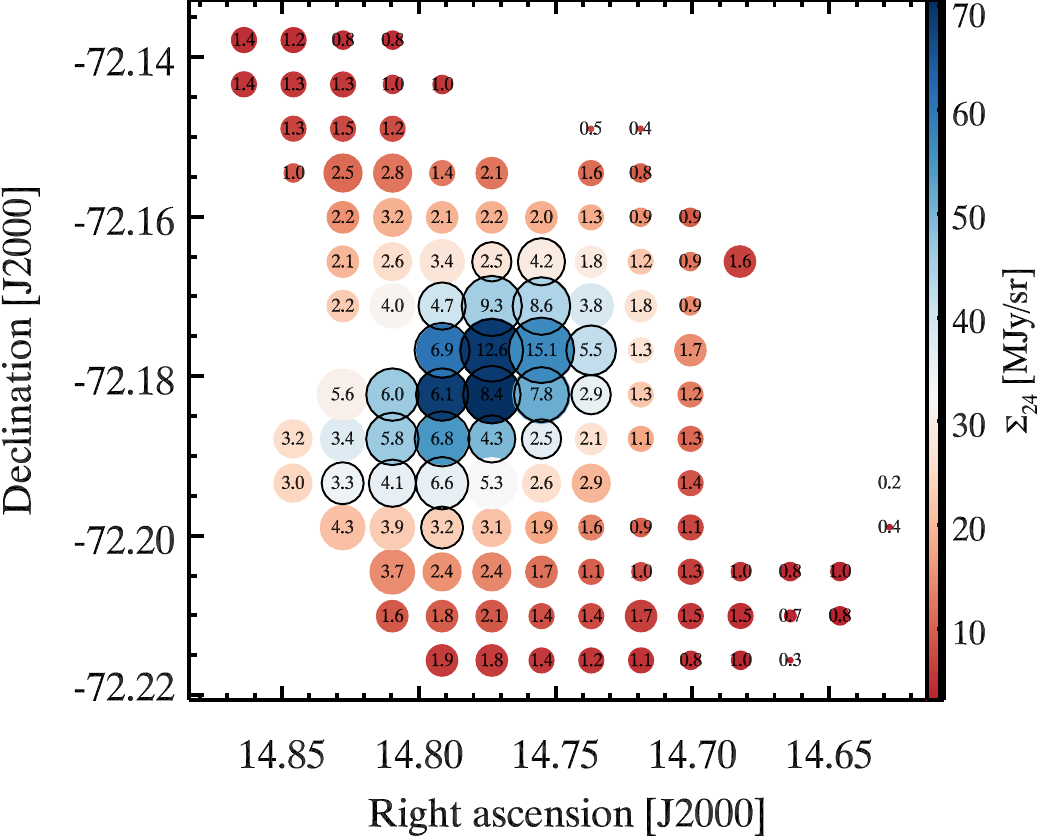}
  \caption{Map of the \ngc which shows the derived stellar mass
    fraction. The size of each point is dependent on \fracmstar and
    its colour depends on \sigmatwentyfour. The mass fraction is
    indicated in each point. It is clear from this diagram that the
    stellar mass fraction peaks on the stellar cluster where the dust
    is heated to higher temperatures due to the intense radiation
    field. The warm pixels identified in
    Fig.~\ref{fig:Sigma_dust_Sigma_stars} are encircled. }
  \label{fig:sfemap}
\end{figure}
The observed variations in \sfr at higher dust/gas surface densities
are clearly correlated with other observables, in particular the dust
colour-temperature (here \sigmatwentyfour/\sigmatwofifty [Jy/Jy]).
Perusal of the data shows that those pixels with
\sigmatwentyfour/\sigmatwofifty$>$0.3, i.e. warm dust, are towards the
rich stellar cluster and all exhibit high values in this diagram. The
position of these warm sight-lines is also indicated in
Fig.~\ref{fig:sfemap}. We further show the power law fit to the data
excluding the lines-of-sight with this very warm dust (exponent
1.6$\pm$0.1). The derived exponent of the power law varies
significantly depending on whether we include these sight-lines, which
is an indication that the lines-of-sight towards the main cluster
occupy a particular part of this diagram.
  
It is interesting to note that the rms scatter around the best-fitting
power law is $\pm$0.4dex, This is similar to the rms observed
($\pm$0.3dex) in the SK-relation of integrated `normal' spiral
galaxies \citep{2012ARA&A..50..531K}. Perhaps, some of the differences
in physical conditions that underpin the variations among spiral
galaxies are encountered within the \ngc star forming complex. If this
is indeed the case, then studies such as the one we present here,
offer a rare opportunity to zoom in to specific environments and
determine the parameters that drive such variations.

In the following we discuss the variations in terms of the stellar
mass fraction of young stars
($\fracmstar=\sigmamstar\,(\sigmamstar+\sigmagas$)$^{-1}$) to
facilitate comparison with star cluster formation models, where
\sigmamstar refers to the mass of the PMS+UMS stars. Moreover, we
prefer this terminology over star formation efficiency, which concerns
the time averaged rate with which a large quantity of interstellar
matter (entire complexes) is converted into stars. On the small scales
we probe here, we measure the stars that have formed earlier and the
remaining ISM column density. It is not evident that these stars have
formed from the ISM we detect now along the line-of-sight nor that
they will still be in this location, if and when the next generation
of stars is born.

Across most of the surveyed area (66 per cent of the pixels) \fracmstar is
below 2 per cent, with the most common value being 1 per cent\footnote{1 per cent
  corresponds to a gas consumption time-scale of 5$\cdot$10$^8$ yr,
  given the duration of the star-formation event of 5\,Myr}, see also
Fig.~\ref{fig:sfemap}. However, there is a significant tail to much
higher values with a maximum of 15 per cent. These very high values
correspond to very high stellar surface densities, \emph{not} to low
ISM column densities (see Fig.~\ref{fig:fitmaps}).

\begin{figure}
  \includegraphics[clip,width=8.8cm]{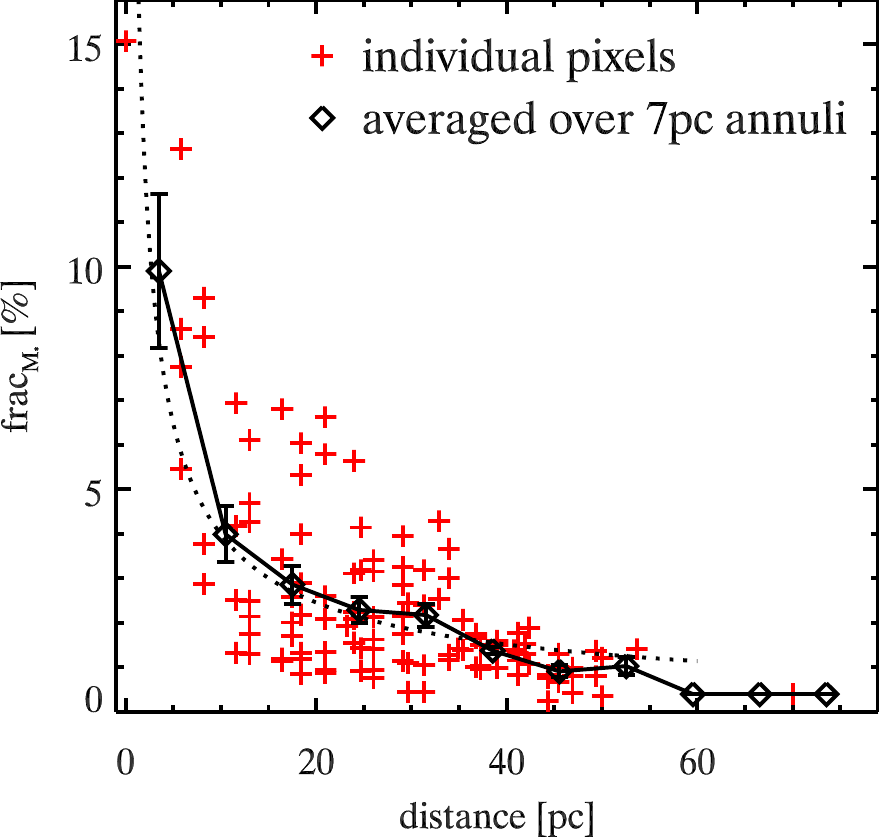}
  \caption{Trend of stellar mass fraction as a function of distance
    from the main cluster (ra=14.755179\degree, dec=-72.171201\degree
    [J2000]). Individual pixel values are shown by red pluses.
    Annular-averaged values (in steps of 7\,pc) are shown with black
    diamonds. The error bars shown are 1$\sigma$ uncertainties on the
    mean. The behaviour can be approximated by \fracmstar [\%] =
    20\%$\cdot$distance [pc]$^{(-0.7)}$ between $\sim$1 and 55\,pc
    (dashed line). }
  \label{fig:star_percent_trend}
\end{figure}

The observed variations in \fracmstar are certainly not random, but
show a definite dependence on other measurable quantities. In
particular, the highest values are found near the main cluster
(Fig.~\ref{fig:sfemap}). We also show the trend of stellar mass
fraction as a function of distance to the main cluster in
Fig.~\ref{fig:star_percent_trend}. Even though there are clear
deviations from spherical symmetry in the region, we observe a
monotonic decrease in \fracmstar with distance which can be
approximated by a power law with exponent -0.7.

\begin{figure*}
  \includegraphics{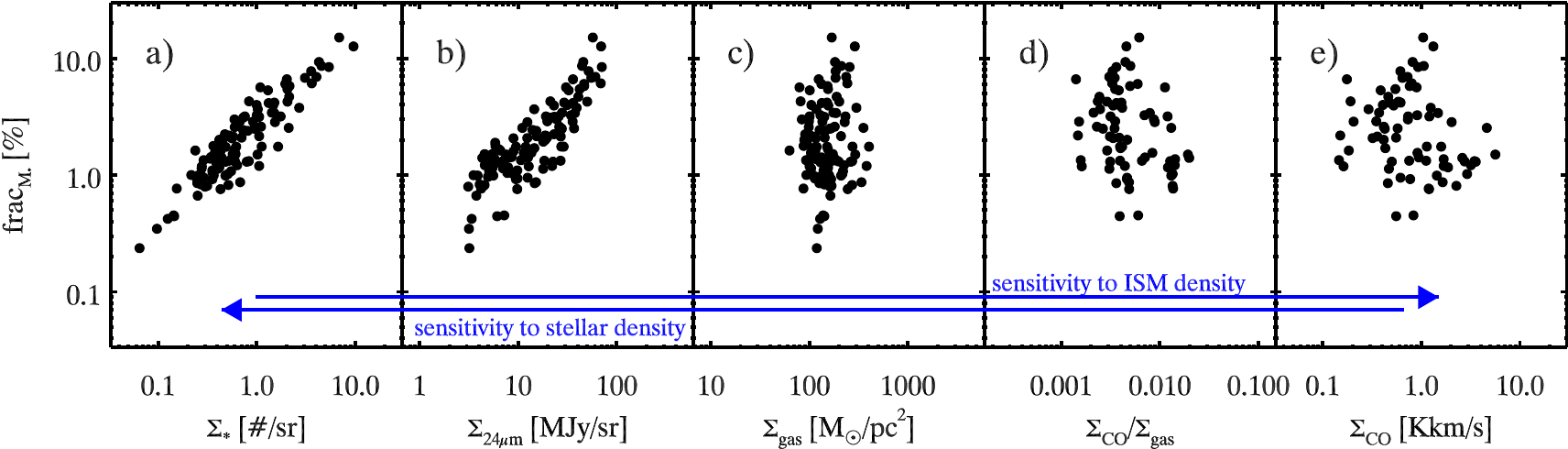}
  \caption{Scatter plots of the stellar mass fraction as a function of
    a number of key tracers. From left to right the tracer measures
    increasingly the ISM density and the presence of dense clumps.
    From right to left the sensitivity to the local stellar density
    increases. \sigmastar is the surface density of the PMS+UMS
    stars. All panels span an identical dynamic range
    ($x_\mathrm{max}/x_\mathrm{min}=y_\mathrm{max}/y_\mathrm{min}=1\,000$).}
  \label{fig:correlations}
\end{figure*}

From the observations and the SED fitting, we have at our disposal a
number of key probes that typify the conditions along the
line-of-sight of each pixel. These range from quantities that measure
directly the stellar densities and the resulting intense radiation
field, to those that depend on both the local radiation field and the
ISM column density and state. In these same ordering, these probes
are: \sigmastar, \sigmatwentyfour, the average radiation field (\uav)
as experienced by the dust grains\footnote{\uav is the average field
  weighted by dust mass and as a result does not measure the ambient
  radiation field if the line-of-sight extinction is significant.},
\sigmagas, the molecular gas fraction and the molecular gas surface
density. Perusal of the data shows that \fracmstar correlates well
with stellar densities and hot dust emission at 24\,$\mu$m
(Fig.~\ref{fig:correlations},a,b). The correlation with the ISM column
density is much less tight (panel c), which is in part due to the fact
that the amplitude of variations in \sigmagas are much smaller that
those is \sigmastar, i.e. the gas map is much flatter than the star
map. When we use \sigmaco/\sigmagas as a proxy for the molecular gas
fraction (panel d), there is no obvious correlation between the
stellar mass fraction and the molecular gas fraction. Specifically,
the highest \emph{and} the lowest stellar mass fractions coincide with
intermediate values of the molecular gas fraction. The same is true
also for the molecular surface density (panel e).
  
The basic conclusion we can draw is that the highest stellar mass
fractions are found where the highest concentrations of stars are
found, towards the central cluster. \emph{There are many stars where
  there are many stars.} We find no obvious parameters of the
current ISM towards the cluster that explain why the cluster formed
there. It is neither the densest nor the most molecular ISM within the
complex. Probably the strong radiation over the past few Myr has
already erased the memory of the conditions that led to the formation
of the cluster.

It should be noted that the central location of the main cluster may
be a clue to why it formed there. It truely is central to the complex
in most tracers. In the ISM column density tracers, in particular, the
main cluster location appears as the intersecting `hub' of three
peripheral dense `arms' (Fig.~\ref{fig:overview} and
\ref{fig:gallery2}). If the main cluster is indeed located at the
meeting point of several dense arms, it possible that the star
formation process was/is being fed by these structures. Whether this
is the case cannot be assessed without kinematic measurements of the
dense gas.

\section{Possible biases}
\label{sec:possible-biases}
We find that the stellar mass fraction is highest towards the rich
central cluster. There are several effects and biases that could
influence this interpretation. 

\subsection{Completeness}
\label{sec:completeness}
The derived relation between the newly formed stars and the ISM
depends on the completeness of the \hst stellar catalogue. The average
completeness for stars which are too faint and crowding has been
accounted for by integrating the completeness corrected mass function
\citep[][see also Sect.~\ref{sec:observations}]{2008AJ....135..173S}.
If there are remaining incompleteness issues in the catalogue, the
\fracmstar would be even higher that we derive now. The observed
variations in \fracmstar could be affected, if there are
large spatial variations in completeness. \\
{\it Crowding:} The regions which could be affected by crowding more
than the average are the regions of the highest stellar density.
However, these regions already have the highest mass fraction of
stars, and an additional correction would place these regions
farther from the mean rather than closer.\\
{\it Extinction:} \ngc exhibits some sight-lines with significant
amounts of dust (Fig.~\ref{fig:gallery2}) which could hide embedded
clusters of young stars. This could affect, in particular, the regions
labelled \#3 and \#4. Both regions have a lower than average stellar
mass fraction and therefore additional embedded young stars would
bring them closer to the average. However, the data argue against a
significant population of embedded sources. First, because the peaks
in the \sigmadust map do correspond to local peaks in the \hst
detected \sigmastar map. Secondly, because there are no corresponding
peaks in the MIPS~24\,$\mu$m maps at these locations, while an
embedded cluster would produce 24~$\mu$m emission. This emission would
be detectable unimpeded by line-of-sight extinction. Finally, these
regions have low observed dust temperatures, while embedded heating
sources would increase their temperature.

\subsection{Variations in gas-to-dust mass-ratio}
\label{sec:variations-gdr}
If there are significant and systematic variation in \gdr, the
conversion from \sigmadust to \sigmagas would be affected.
Specifically, if the dust in the central part has been mostly
destroyed, we would grossly underestimate the \sigmagas there, which
would bring these points closer to the general trend (to the right in
Fig.~\ref{fig:Sigma_dust_Sigma_stars}). The SED modelling shows that
these regions are extremely depleted in PAHs. Is it plausible that the
dust has also been destroyed? The increase required in the assumed
value of \gdr for the central region of \ngc in order to make the
variations in \fracmstar disappear is a factor of 10 (\gdr equal to
12\,500 instead of 1250).
  
An argument against significant dust destruction (or more generally
significant variations in \gdr) is the very similar behaviour of
\halpha and TIR (and 24\,$\mu$m) emission. The rms of
\sigmatir/\sigmahalpha of the 115 pixels across the whole region is
only 20 per cent and its behaviour is not spatially correlated with the
stellar density or the radiation field density. This similar behaviour
is also reflected in Fig.~\ref{fig:sfr_comparison}\footnote{The
  scatter in \sigmatir/\sigmahalpha of 20 per cent is much smaller than the
  scatter -- about the mean trend in Fig.~\ref{fig:sfr_comparison} --
  in the \sigmatir and \sigmahalpha, individually. This indicates that
  an above (below) average value of \sigmahalpha corresponds to an
  above (below) average value in \sigmatir.}. For the relatively low
extinction conditions in \ngc, \sigmahalpha measures the chance for
the starlight to be absorbed locally by the gas and \sigmatir measures
the same for the dust, the very similar behaviour (on scales of
6$\times$6\,pc$^{2}$) of these tracers suggest a similar
proportion of dust to gas across the region we studied. In case of
very reduced dust abundances towards the central region, one would
expect a limb-brightened shell in the dust emission, contrary to what
is observed (Fig.~\ref{fig:fitmaps}a).

A relatively constant \gdr is further supported by the analysis of
\cite{2008ApJ...680..398L} who measured the gas phase abundances along
13 different sight-lines across the \ngc complex. They find only minor
variations in the gas phase abundances. Most importantly they find
only 0.1dex variations in the S/H ratio while sulphur can be depleted
into dust grains and strongly varying dust abundances would lead to
variations of the abundance of sulphur remaining in the gas phase. We
conclude that variations in \gdr that could explain the variations in
\fracmstar are improbable.

\section{Discussion}
\label{sec:disc--concl}
\subsection{Effects of evolution}
\label{sec:effects-evolution}
Our observations provide a snapshot of \ngc at one instance, roughly
3$-$5\,Myr after the onset of the main star-formation event. During
the star-formation epoch the amount of gas transformed into stars
\emph{must} evolve. Is the observed scatter in \fracmstar caused by
studying only this snapshot? In other words, is it possible that after
the star formation has ceased, the \fracmstar will be more
homogeneous? Within the complex, different regimes can be discerned,
as is illustrated in Fig.~\ref{fig:gallery2}.
\begin{itemize}
\item {\it Region \#1:} Main cluster with high dust surface densities
  and high number of stars. The region is detected in CO with
  moderate brightness.
\item {\it Region \#2:} Secondary cluster with fewer stars and less
  dust. Appears to be a `scaled-down' version of the main cluster
  with bright \halpha and MIPS~\,24\,$\mu$m emission and a
  moderately high \fracmstar.
\item {\it Region \#3:} This blob to the north-west of the complex has
  the highest dust surface density and low measured \fracmstar. It is
  detected in CO but the estimated molecular gas fraction is low.
\item {\it Region \#4:} The molecular `spur' to the north-east has a
  small stellar mass fraction, high dust surface an no associated
  \halpha emission.
\end{itemize}
It is unlikely that these regions will all produce the same stellar
mass fraction in the end. It can be argued that region \#4, with its
low current \fracmstar but large reservoir of molecular gas, has the
potential to still form a significant number of stars and thus boost
its final mass fraction. However, the conditions in region \#1$-$\#3
do not seem to favour prominent continued star formation. Their ISM
seems lacerated with ionized gas, as testified by the prominent
\halpha emission and based on the CO map the molecular gas fraction is
also low. We assess that, most likely, the observed differences
between regions \#1$-$\#3 will persist.

\subsection{Gas removal and stellar motions}
\label{sec:gas-removal}
We find a clear trend in stellar mass fraction from the central parts
of \ngc to its outskirts. Within $\sim$15\,pc \fracmstar$\approx$10
per cent while between 15-50\,pc it is $\sim$2 per cent. These zones
are dominated by the cluster and the dispersed young stellar
population, respectively as shown with the auto-correlation analysis
of the stellar distribution \citep{2014MNRAS.439.3775G}. The observed
decrease of \fracmstar with distance from the cluster is a signpost of
lower \sfe on the periphery (dispersed SF) than in the densest central
parts (clustered SF). However, there are two caveats to consider, both
related to the relative displacement of the stars and the ISM after
birth.

\begin{figure}
  \includegraphics[clip,width=8.8cm]{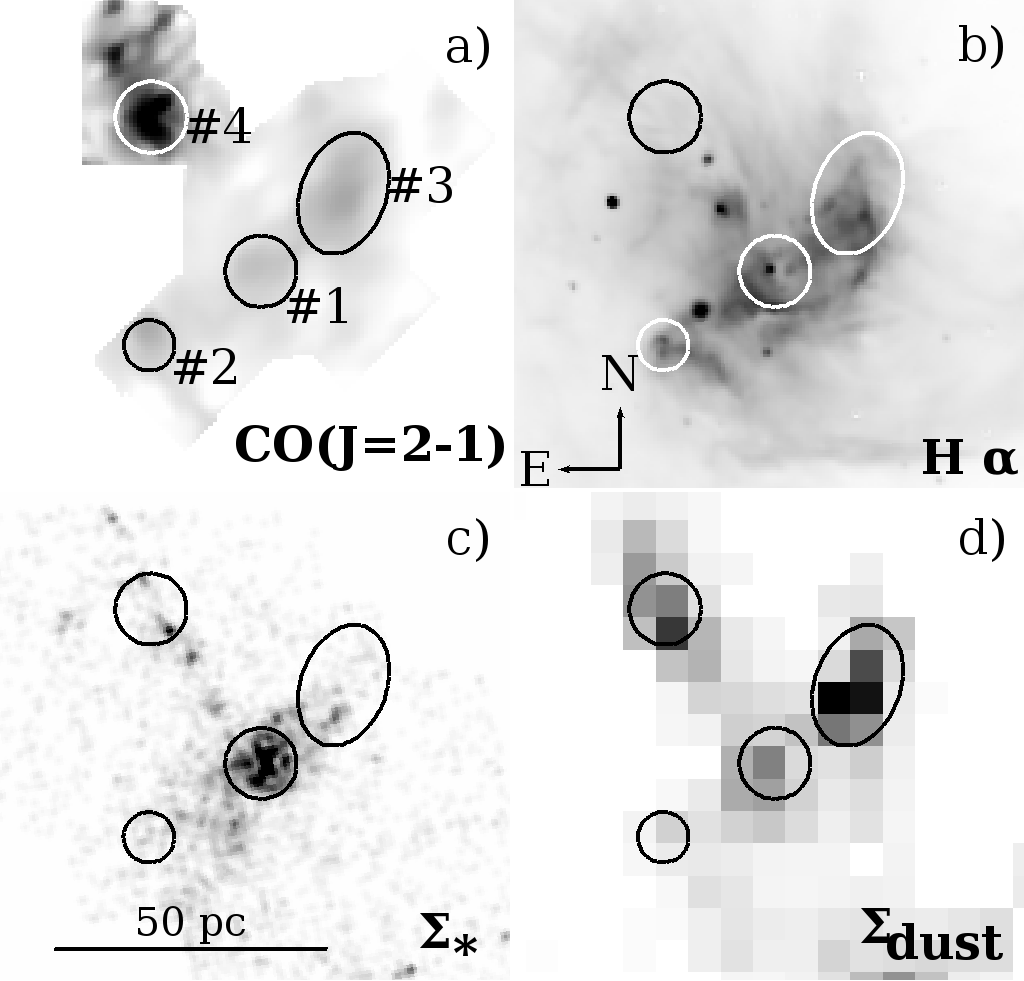}
  \caption{Distinct regions within the \ngc complex as observed in a
    number of key tracers: CO(J=2-1) integrate intensity map
    \citep[][panel a]{2000A&A...359.1139R}, \halpha emission
    \citep[][panel b]{2000ASPC..221...83S}, young star density map
    based on \citet[][panel c]{2006ApJS..166..549G} and the dust
    surface density (this work, panel d). All maps are shown in linear
    scale. We indicate four regions corresponding to local maxima in
    CO emission.}
  \label{fig:gallery2}
\end{figure}

A naive explanation for the present day systematic behaviour of the
stellar mass fraction is the accumulation of stars, born in the
periphery, in the central cluster. If such a process was possible, it
would boost the observed \fracmstar in the cluster while reducing its
value outside. However this is an implausible scenario. Consider the
free-fall time (\tauff) of a star born in the periphery to be pulled
into the cluster by gravity:
\begin{equation}
  \label{eq:1}
  \tauff = \frac{\pi}{2} \frac{d^{3/2}}{\sqrt{G M_{cl}}},
\end{equation}
where $d$ is the distance from the cluster, $M_{cl}$ is the mass of
the cluster (gas plus stars) and $G$ is the gravitational constant.
\tauff is the \emph{shortest} time-scale for a star to migrate towards
the cluster, ignoring friction, virial velocities, rotational motions
and the existence of any matter outside of the cluster that will
exercise gravitational pull. For the observed $M_{cl}$ of
$\sim$10$^5$\,\msun \newline
($=N_{\star < 15pc}\mcat\,\fracmstar^{-1}$ $=2300\cdot4.3/0.1$) and a
distance of 30\,pc (the distance between region \#1 and \#3 in
Fig.~\ref{fig:gallery2}), \tauff$\approx$9\,Myr, considerably longer
than the ages of the PMS stars.

The second effect that could alter the observed \fracmstar over time,
is gas removal from the central region. When a cluster is formed, the
gas left over from the star formation process may be rapidly removed
by stellar winds \citep{1984ApJ...285..141L}. More recent work,
however, suggests that the winds (and later, supernovae) may escape
through low density channels formed in a hierarchical ISM
\citep[e.g.][]{2013MNRAS.431.1337R,2014MNRAS.442..694D} causing much
of the residual gas to remain. This seems to be the case for \ngc.
\hst images of \ngc in the light of \halpha show the presence of
residual gas. Molecular gas also appears to be present in the centre
of the cluster despite the proximity of O stars
(Fig.~\ref{fig:gallery2}). \chandra observations reveal a low diffuse
X-ray flux \citep{2002ApJ...580..225N}, which indicates either that a
supernova explosion has yet to occur, or the hot gas has escaped the
region. Therefore gas expulsion cannot have produced the position
dependence of \fracmstar. In further support of this conclusion are
the findings of \citet[][]{2012ASPC..465..416S} and Smith et al. (in
prep). These authors present high spectral resolution observations of
\halpha and [O {\sc iii}] to examine the ionized gas motions in \ngc.
They find that the gas is quiescent. Data over ten slit positions
covering the ionized nebula show no evidence for velocity splitting,
even at the centre of the cluster. These results indicate that the gas
is not being expelled by stellar winds but rather the dominant form of
interaction is via stellar radiation to produce a classical expanding
H~{\sc ii} region. 

An additional argument against the importance of efficient gas removal
leading to a cavity in the central part of our map comes from the
\spitzer detected embedded YSO candidates
\citep{2007ApJ...669..327S,2013ApJ...778...15S}. These sources are
young ($\lsim$1 Myr) and several of them are detected towards the
central cluster. The number of detected YSO candidates towards the
centre of \ngc is probably a lower limit, set by the difficulty of
detecting individual point-sources which exhibit a mid-IR excess in
the crowded cluster region against the strong extended emission
component due to the hot dust. The \sfr derived from these objects
\citep[$>$3.2\,10$^{-3}$\,\msun\,yr$^{-1}$,][]{2007ApJ...669..327S} is
fully compatible with the average \sfr for the entire 5\,Myr event. If
the clearing of the natal gas had taken place, one would expect to
find few of these very young sources towards the central part of our
map and probably also a diminishing trend of the star formation rate.

We conclude that the difference of a factor of $\sim$5 in observed
\fracmstar, most likely reflects a real difference in the average \sfe
in the clustered versus the dispersed star-forming regimes. This
finding is consistent with theoretical models that assume that the
star-formation efficiency is proportional to the free-fall time, hence
denser regions have higher efficiencies
\citep[e.g.][]{2000ApJ...530..277E,2007ApJ...654..304K,2012MNRAS.426.3008K,2013MNRAS.428.1950H}.

\subsection{High SFE -- linked to low metallicity?}
The \sfe of \ngc is high even after averaging over $\sim$150\,pc
(radius) scales (Fig.~\ref{fig:Sigma_dust_Sigma_stars_large}). 
As discussed in \citet{2014MNRAS.439.3239K}, this may be explained by
the bias introduced by having chosen an active star forming complex as
the object of study. These authors predict an offset of a factor of
$\sim$3 for the size of the regions we study , similar to what is
observed.
Additionally, this high value may be related to an above average
star-formation efficiency in the clustered environment which dominates
\ngc.
Nevertheless, it is still interesting to discuss the possible role of
the lower metallicity for the following reasons. {\it i)}
\citet{2008ASPC..390..149K} shows that low-metallicity galaxies like
the SMC, as a whole, lie a bit (factor of $\sim$3) above the general
SK-relation. {\it ii)} The largest radius we measured is 150\,pc. It
is not obvious that by averaging to a larger distance, the derived
\sfe will decrease significantly -- as it is also not obvious that it
will stay high -- because the next prominent star-forming complex (LHA
120-N76) is only about 300\,pc projected distance away from \ngc.

Because the \sfe within individual star-forming clouds is a critical
parameter for theories of galaxy evolution and galactic-scale star
formation, with observable consequences on for example the galaxy
luminosity function, there are many recent studies that have tried to
assess whether the conversion of gas to stars depends on local ISM
conditions and/or global galaxy parameters.
\citet{2011MNRAS.415...61S} investigated the global gas content and
star formation rates in $\sim$200 galaxies in the redshift range
between 0.025 and 0.05, finding a clear decrease in the star formation
efficiency with increasing stellar mass. Resolved studies of nearby
galaxies, meanwhile, have shown that the star formation efficiency on
$\sim$kpc scales varies with several ISM and Galactic-scale
parameters, including stellar mass, dust-to-gas ratio, rotation
velocity, orbital time and Hubble type
\citep[e.g.][]{2013AJ....146...19L}.

While the influence of any one parameter is difficult to isolate, a
general conclusion that can be drawn from these studies is that low
mass, low metallicity, late-type galaxies appear to be more efficient
at forming stars from their molecular reservoir than high mass,
metal-rich, early-type disc-galaxies. This could be in part a direct
effect of metallicity: \citet{2011ApJ...737L..20D} show that low
metallicity star forming regions may lie above the SK-relation for
normal metallicity galaxies by a factor of 0.3dex as a consequence of
higher \sfe caused by less effective removal of the gas from the star
forming region by stellar winds at low metallicity
\citep{2011MNRAS.415.3439D}. The observed trend may also be due to an
increasing fraction of molecular material located outside bound
star-forming structures as the stellar mass and hence ISM pressure
increases, although other effects are likely to contribute to the
observed trends.

Our results in Sect.~\ref{sec:global-comparison} are consistent with
the picture emerging from lower resolution studies that molecular gas
in low-metallicity, low-mass systems is somewhat more efficient at
forming stars. In our case, we find that the mass of young stars in
\ngc is high relative to its gas content, even after we average our
measurements over scales that approach the spatial resolution of
nearby galaxy surveys. However, we caution the reader to realise that
the presented analysis of \ngc alone does not allow us to derive the
global SMC star-formation efficiency nor conclude that it is above
average.

\section{Summary}
We present a detailed comparison of the surface density of young
(PMS+UMS) stars (\sigmastar) and dust surface density (\sigmadust)
across the face of the most prominent star forming complex \ngc(N66)
in the SMC. \sigmastar is derived from individual stars identified in
the colour-magnitude diagram based on \hst photometry. \sigmadust is
obtained via radiative transfer modeling of the infrared to
submm spectral energy distribution. We measure these
quantities in 115 independent pixels of 6$\times$6\,pc$^{2}$ out to
distances of $\sim$50\,pc from the main cluster, covering an area of
$>$4000\,pc$^2$.

We find a correlation between \sigmastar as a function of \sigmadust
with a considerable scatter. A power law fit to the data yields a
steep relation with an exponent of 2.6$\pm$0.2. We find that
sight-lines towards the central $\lsim$15\,pc exhibit
\emph{systematically} high values of \sigmastar/\sigmadust by a factor
of $\sim$5 compared to the rest of the complex.

We convert \sigmastar to average star formation rate (\sfr) surface
densities (\sigmasfr in \msun\,yr$^{-1}$\,pc$^{-2}$) using {\it 1})
the observed mass function to derive the total stellar mass formed and
{\it 2}) the time span over which these young stars have formed
(5\,Myr). This \sigmasfr is much less biased to the contribution of
high-mass stars than luminosity based \sfr tracers. The derived total
\sfr (4$\pm$1$\cdot$10$^{-3}$ \msun\,yr$^{-1}$) is consistent (within
10 per cent) with \sigmasfr estimated from the extinction corrected
\halpha emission integrated over the entire \halpha nebula.

On small scales (6$\times$6\,pc$^{2}$) the \sfr estimated from
indirect tracers (ionized gas or dust emission) compares very poorly
with the \sfr based on star counts. Across most of the studied region,
\halpha systematically underestimates the local \sfr, reaching a
factor of 10 discrepancy in the central part of the complex. This
discrepancy is due to ionizing photons escaping the local volume where
the young stars are counted.

We convert \sigmadust to gas surface density (\sigmagas) using a
constant gas-to-dust mass ratio (\gdr = 1250). This value is
calibrated on our dust SED fitting method to yield the measured gas
mass (from HI and CO) on a large scale, within a 500\,arcsec radius
centred on \ngc. On this scale most of the gas mass is contained in
the atomic phase and the influence the uncertain conversion factor
from CO intensity to H$_2$ column density is small.

Using the derived \sigmagas and \sigmasfr, we place the pixels on a
Schmidt-Kennicutt (SK) diagram. We find that individual pixels fall
systematically above the SK-relation for integrated disc-galaxies by
on average a factor of $\sim$7. This offset decreases as larger beams
are used, consistent with the prediction from
\citet{2014MNRAS.439.3239K}.

The observed offset is reminiscent of the results of
\citet{2010ApJ...723.1019H} and \citet{2010ApJS..188..313W} who show
that the measured \sfr lies above the galaxy averages when
`zooming-in' to parsec-scale on star forming clumps in the Milky Way.
This is probably caused by less dense gas, which is inefficient in
forming stars, that is included in the galaxy-scale averages but not
included when `zooming-in' on individual clumps.
A main difference with our analysis is that, even
though we analyse 6$\times$6\,pc$^{2}$ sized regions, we cover the
full star-forming complex and find that the entire active region
($\sim$50\,pc) lies consistently above the relation. In other words
for \ngc one needs to `zoom-out' beyond 50\,pc to probe this less
dense, non-star forming gas. The average over a larger area (90\,pc
radius) lies closer to the SK-relation but remains high by a factor of
$\sim$3.

We present a map of the observed stellar mass fraction (\fracmstar =
\sigmastar(\sigmagas+\sigmastar)$^{-1}$). \fracmstar is systematically
high ($\sim$10 per cent) within the central 15\,pc of the complex and
systematically lower outside (2 per cent). We interpret this
difference as the results of a higher star formation efficiency in the
central zone. The inner 15\,pc is dominated by young stars belonging
to a centrally condensed cluster, while the outer parts are dominated
by a dispersed population \citep{2014MNRAS.439.3775G}. Therefore, the
observed behaviour could reflect a change in star-formation efficiency
between clustered and non-clustered star-formation.
 
\vspace{-0.8cm}
\section*{Acknowledgements}
Herschel is an ESA space observatory with science instruments provided
by European-led Principal Investigator consortia and with important
participation from NASA.
Based in part on observations made with the NASA/ESA Hubble Space
Telescope, obtained from the data archive at the Space Telescope
Science Institute (STScI). STScI is operated by the Association of
Universities for Research in Astronomy, Inc. under NASA contract NAS
5-26555.
This work is based in part on observations made with the Spitzer Space
Telescope, which is operated by the Jet Propulsion Laboratory,
California Institute of Technology under a contract with NASA.
This publication is based in part on data acquired with the Atacama
Pathfinder Experiment (APEX). APEX is a collaboration between the
Max-Planck-Institut fur Radioastronomie, the European Southern
Observatory, and the Onsala Space Observatory.
We thank Sean Points and Roger Leighton for kindly making available
the MCELS \halpha data.
SH and RSK acknowledge support from the {\em Deutsche
  Forschungsgemeinschaft} (DFG) in the collaborative research project
SFB881 “The Milky Way System” (subprojects B1, B2, and B5) as well as
in the priority program SPP 1573 “Physics of the Interstellar Medium”.
RSK furthermore acknowledges support from the European Research
Council under the European Community’s Seventh Framework Programme
(FP7/2007-2013) via the ERC Advanced Grant STARLIGHT (project number
339177).
DAG kindly acknowledges financial support by the German Research
Foundation through grant GO 1659/3-1.
SD is supported by a Marie-Curie Intra European Fellowship under the
European Community's Seventh Framework Program FP7/2007-2013 grant
agreement no 627008
AH acknowledges support from the Centre Nationl d'Etudes Spatiales
(CNES) and funding from the Deutsche Forschungsgemeinschaft (DFG) via
grants SCHI 536/5-1 and SCHI 536/7-1 as part of the priority program
SPP 1573 'ISM-SPP: Physics of the Interstellar Medium'.
MYL acknowledges supports from the DIM ACAV of the Region Ile de
France.

\vspace{-0.8cm}
\bibliographystyle{mn2e}
\bibliography{hony_ngc346}
\label{lastpage}
\end{document}